\documentclass[a4paper,11pt]{article}
\pdfoutput=1
\usepackage{jcappub}

\usepackage{epsfig,amsmath,amssymb,bm,color,graphicx}
\usepackage{afterpage}
\usepackage{pdfpages}
\usepackage{subfig}

\graphicspath{{./figures/}}

\newcommand{\lya}{Lyman-$\alpha$}
\newcommand{\mlya}{\mathrm{Ly}\alpha}

\newcommand{\Mpch}{h^{-1}\mathrm{Mpc}}
\newcommand{\cMpch}{c\mathrm{Mpc}/h}
\newcommand{\hMpc}{h\,\mathrm{Mpc^{-1}}}
\newcommand{\skm}{\mathrm{s\,km^{-1}}}

\newcommand{\vN}{{\vec N}}
\newcommand{\vx}{{\vec x}}
\newcommand{\HI}{H{\sc ~i}}

\newcommand{\be}{\begin{equation}}
\newcommand{\ee}{\end{equation}}

\usepackage{color}

\begin{document}

\title{Absorber Model: the Halo-like model for the Lyman-$\alpha$ forest}

\author[a]{Vid Ir\v{s}i\v{c},}
\author[a]{Matthew McQuinn}

\affiliation[a]{University of Washington, Department of Astronomy, 3910 15th Ave
NE, WA 98195-1580 Seattle, USA}

\emailAdd{irsic@uw.edu}

\date{\today}

\abstract{ We present a semi-analytic model for the \lya\ forest that is
  inspired by the Halo Model.  This model is built on the absorption line
  decomposition of the forest.  Flux correlations are decomposed into those
  within each absorption line (the 1-absorber term) and those between separate
  lines (the 2-absorber term), treating the lines as biased tracers of the
  underlying matter fluctuations.  While the nonlinear exponential mapping
  between optical depth and flux requires an infinite series of moments to
  calculate any statistic, we show that this series can be re-summed (truncating
  at the desired order in the linear matter overdensity).  We focus on the
  $z=2-3$ line-of-sight power spectrum.  Our model finds that 1-absorber term
  dominates the power on all scales, with most of its contribution coming from
  \HI\ columns of $10^{14}-10^{15}\;\mathrm{cm^{-2}}$, while the smaller
  2-absorber contribution comes from lower columns that trace overdensities of a
  few.  The prominence of the 1-absorber correlations indicates that the
  line-of-sight power spectrum is shaped principally by the lines' number
  densities and their absorption profiles, with correlations between lines
  contributing to a lesser extent.  We present intuitive formulae for the
  effective optical depth as well as the large-scale limits of 1-absorber and
  2-absorber terms, which simplify to integrals over the \HI\ column density
  distribution with different equivalent-width weightings.  With minimalist
  models for the bias of absorption systems and their peculiar velocity
  broadening, our model predicts values for the density bias and velocity
  gradient bias that are consistent with those found in simulations.  }

\maketitle

\section{Introduction}

The Ly$\alpha$ forest is one of the primary tools for understanding the
intergalactic medium as well as the Universe's initial conditions and expansion
history.  It has been used to constrain cosmological parameters
\citep{mcdonald00,zaldarriaga01,croft02,zaldarriaga03,seljak03,mcdonald03,viel04,viel04bis,viel04hrwmap,mcdonald05,mcdonald06,seljak06,slosar11,busca13,slosar13,palanque13,palanque15,bautista15,baur17,bautista17,bourboux17},
the temperature and photoionization rate of the intergalactic gas
\citep{schaye00,ricotti00,theuns00,mcdonald00,viel06,theuns02,bolton08,lidz10,bolton10,becker11,garzilli12,rudie12,lee14,boera14,bolton14,rorai17,hiss17},
and dark matter models
\citep{narayanan00,viel05,uros06,viel08,bird10,viel13WDM,baur15,yeche17,irsic17a,
  irsic17b,armengaud17}. However, the Ly$\alpha$ forest is sufficiently
nonlinear that perturbative methods cannot describe many of the spatial scales
used for these constraints.  Cosmologists' understanding of how the Ly$\alpha$
forest traces the large-scale density and velocity fields (as well as how it is
shaped by both cosmological and astrophysical parameters) derives primarily from
running suites of cosmological hydrodynamic simulations
\citep[e.g.][]{mcdonald03, bolton16}.

In the pursuit of a new tool for understanding the Ly$\alpha$ forest, this paper
develops a semi-analytic model that is inspired by the Halo Model.  Of all
analytic large-scale structure models, the Halo Model has met the most success
at bridging linear and nonlinear scales for many tracers of the large-scale
cosmic matter field \citep{ma00, seljak00, cooray02}.  The key insight of the
Halo Model is that small-scale correlations are dominated by the clustering
within individual halos (the one-halo term) and large-scale correlations by the
clustering of distinct halos (the two-halo term).  The former correlations are
Poissonian and can be related to the tracers' profiles within halos, whereas the
simplest implementations of the latter halo-halo correlations use linear-order
cosmological perturbation theory.  Intuitive formulae for the clustering of
cosmological objects result from the sum of these correlations.  While there are
issues with the description of mildly nonlinear scales and with the level of
shot noise \citep[e.g.][]{seljak09, mead15, baldauf13, ginzburg17}, the Halo
Model has provided a platform for understanding nonlinear structure formation,
the clustering of galaxies of different types, and the anisotropies in various
radiation backgrounds \citep[for a review see][]{cooray02}.

While our model is inspired by the Halo Model, in contrast to the Halo Model,
most of the gas in the Ly$\alpha$ forest is not bound to dark matter halos.
Rather the forest primarily traces the voids, sheets and filaments smoothed on
the gas' Jeans scale \citep[for recent reviews see][]{meiksin09, mcquinn15}.
Numerical simulations of the forest show that these voids, sheets, and filaments
manifest as Ly$\alpha$ absorption lines with various neutral hydrogen columns
\citep{miralda96, hernquist96}.  Indeed, many studies of the Ly$\alpha$ forest
have developed an understanding for the properties of these absorption lines --
their linewidths, columns, and frequencies
\citep{schaye00,ricotti00,prochaska10,rudie12,sdss12gas,bolton14,hiss17}.
(There is also a theoretically-motivated relation that relates their column and
density \citep{schaye01}, which we use to model the clustering of lines.)  The
halo-like model developed here uses the absorption lines as the stochastic
element rather than halos, breaking Ly$\alpha$ forest statistics into
correlations within individual absorption lines and between lines. Much like in
the Halo Model, intuitive expressions lead to a result that is a function of
these properties and the linear matter density. Similar difficulties arise as in
the Halo Model in this model's treatment of mildly nonlinear scales.

The `Absorber Model' presented here is not the first semi-analytic model for the
Ly$\alpha$ forest.  Before the revolution in our understanding of the forest
that occurred with the first hydrodynamic cosmological simulations
\citep{miralda96, hernquist96}, Rees \cite{rees86} developed a halo-like model
in which the absorption is from bound gas within halos; this paper updates this
approach to the modern picture in which the Ly$\alpha$ forest predominantly
traces more diffuse structures.  Models based on linear perturbation theory have
also been developed, but generally fail to explain the properties of the
Ly$\alpha$ forest \citep{bi97}.  Models that map the linear-theory density field
into a nonlinear one via a lognormal transformation have been more successful
\citep{bi97} and to date are the most used semi-analytic model, often to
generate mock forest spectra \citep{font12,bautista17}.  However, as lognormal
models rely on an ad hoc transformation, they provide less intuition into how
different components of the forest contribute than in the model presented here.

This paper is organized as follows.  Section~\ref{sec:absorber_model} presents
the model, Section~\ref{sec:implementation} describes how we implement the model
(the linewidths, column density distributions, and bias parameters) as well as
the simulations we use for comparison, and Section~\ref{sec:main} analyzes the
different effects that shape the power spectrum in the simulations and
especially in the model.  We discuss key takeaways in
Section~\ref{sec:conclusion}.  Throughout we assume a $\Lambda$CDM cosmology
with $\Omega_m = 0.308$, $\Omega_\Lambda = 0.692$, $h=0.678$, $n_s = 0.961$ and
$\sigma_8 = 0.829$.

\section{Absorber model}
\label{sec:absorber_model}

Our approach follows that taken for the Halo Model, except for two important
distinctions.  First, instead of modeling the Ly$\alpha$ forest as being composed of dark
matter halos, we model it as being composed of discrete
absorption lines each with some optical depth profile.  The standard halo
model calculation translates over naturally when considering
correlations in the optical depth field.  Second, the observable in the forest
is \emph{not} the optical depth $\tau$, but the normalized flux, $\exp[-\tau]$.
Even calculating the mean of $\exp[-\tau]$ requires computing an infinite series
of moments in $\tau$, in contrast to the quadratic order that many Halo Model
calculations require.  We do this computation here, showing fortunately that
this series can be re-summed into a compact form.  A final remark before we
introduce the model, and in analogy to the Halo Model (where the profile of the
chosen tracer within a dark matter halo is usually characterized by just the
halo's mass), we assume that an absorber's optical depth profile depends on the
\HI\ column density of lines, $N_{\rm HI}$. For simplicity, we will use the symbol $N$ to
denote the hydrogen column density through this section. As the Halo Model can be extended to
account for other variables such as the distribution of halo concentrations, our
model can be easily extended to include other properties that shape the absorber
optical depth profile.

To begin, we can write the optical depth at position $x$ along a sightline as a
sum over the optical depth profile from all absorbers 

\be 
\tau(x) = \sum_{i } p_i \tau_{\mathrm{a}}(x - x_i|N_{i}),
\label{eq:taux}
\ee 
where $p_i$ is the probability for an absorber with neutral hydrogen column
density $N_{i}$ to be at position $x_i$. The sum goes over all possible positions and column densities, where
we have discretized both quantities into bins with width $\Delta N$ and $\Delta x$ that are
chosen to be sufficiently small so that each $p_i$ is either zero or one. (We use
one index to enumerate all possibilities in both quantities to simplify
notation.) Finally,

\be 
\tau_{\mathrm{a}}(x | N) \equiv
\sigma_0 N W(x|N),
\label{eq:profile}
\ee
is the optical depth profile of an absorber, where $\sigma_0$ is the
velocity-integrated cross section (we use velocity units for $x$)\footnote{The
  cross section is a constant for a given absorber, defined in terms of
  universal constants as $\sigma_0 = \frac{\alpha_{fs} h}{2m_e c}
  \lambda_{\mlya} f_{\mlya}$, where $\alpha_{fs}$ is the fine structure
  constant, $h$ is Planck constant, $m_e$ is electron mass and $c$ is the speed
  of light. The only thing that changes between one absorber and the other is
  the transmission properties, e.g. for \lya\ transmission we have
  $\lambda_{\mlya}$ as the wavelength of the \lya\ transmission and $f_{\mlya}$
  is the oscillator strength of the transmission. Definition of $\sigma_0$ as
  above assumes that position $x$ is given in velocity units. If $x$ were
  instead in distance units, the definition of $\sigma_0$ acquires additional
  factor of $(1+z)/H(z)$.}, and $W(x|N)$ is a line-profile function with unit
norm.  For example, for thermal Doppler broadening plus natural broadening,
$W(x|N)$ is given by a Voigt profile.

We can rewrite Eq.~\ref{eq:taux} for the optical depth as 
\be
\tau(x) = \int dN_1 \int dx_1 \tau_1(x) \sum_i p_i \,
\delta_D(N_1 - N_i) \delta_D(x_1 - x_i),
\label{eq:taux2}
\ee 
where we have introduced two Dirac $\delta$-functions ($\delta_D$) and have
simplified notation to $\tau_1(x) \equiv \tau_a(x-x_1|N_1)$.

To evaluate the moments of the flux field, $F = \exp\left[ -\tau \right]$, we
make use of the cumulant theorem, which for the first moment, the mean normalized flux, yields: 
\be
\langle F \rangle \equiv \langle e^{-\tau}\rangle 
 = \exp\left[-\langle \tau \rangle_c
  + \frac{1}{2}\langle \tau^2 \rangle_c - \frac{1}{3!}\langle \tau^3
  \rangle_c + \cdots\right],
  \label{eqn:mF}
\ee
where
\begin{align}
\langle \tau \rangle_c \equiv \langle \tau \rangle \notag; ~~~~\langle \tau^2
\rangle_c \equiv \langle \tau^2 \rangle - \langle \tau \rangle^2 \notag;
~~~~\langle \tau^3 \rangle_c \equiv \langle \tau^3 \rangle - 3\langle \tau^2
\rangle \langle \tau \rangle + 2\langle \tau \rangle^3; ~~~~\cdots
\end{align}
Thus, the moments of field $F$ can be expressed as an infinite sum over the
cumulants (and moments) of the optical depth field $\tau$. In itself this would
not necessarily be an advantage, unless the series converges and can be re-summed,
as we show is the case.

We initially aim to compute the first and second moments of the optical depth.
This computation requires noting that 
\be 
\langle p_i^m \rangle = f(N_i) \Delta N \Delta x, 
\ee 
where $f(N)$ is the \HI\ column density distribution, and that
\be 
\langle p_i p_j\rangle = \langle p_i \rangle \langle p_j \rangle \left[ 1+
  \xi_{ij}(x_i-x_j) \right], 
\ee 
where the absorber correlation function is $\xi_{ij}(x) \equiv \xi(x|N_i,N_j)$.
We take moments of
Eq.~\ref{eq:taux2}, using $\sum \rightarrow \int \frac{dN dx}{\Delta N \Delta
  x}$ to eliminate sums, yielding
\begin{align}
\langle \tau \rangle &= \int dN_1 \int dx_1 f_1 \tau_1; \notag 
\langle \tau^2 \rangle - \langle \tau \rangle^2 &=  \int dN_1 \int dx_1 f_1 \tau_1^2  +
\int_N \int_x d^2\vN d^2\vx\, f_1\tau_1 f_2 \tau_2 \xi_{12}(x_2-x_1),
\end{align}
where we have abbreviated $f_i \equiv f(N_i)$, $\xi_{ij}(x) \equiv \xi(x|N_i,N_j)$, and again $\tau_i \equiv
\tau_a(x - x_i|N_i)$.  We have also introduced the notation of integrals over column
densities and spatial coordinates as $\int_N d^{n}\vN \equiv \int_N dN_1
dN_2 \dots dN_n$ and  $\int_x d^{n}\vx \equiv \int_x dx_1
dx_2 \dots dx_n$, a notation we continue below.

We can do the same exercise for the third moment of the optical depth.  Using
that

\be \langle p_i p_j p_k \rangle = \langle p_i \rangle \langle p_j \rangle
\langle p_k \rangle \left( 1+ \xi_{ij} + \xi_{jk} + \xi_{ki} + \zeta_{ijk}
\right), \ee where $\zeta_{ijk}$ is the absorber three-point function, a bit of
algebra yields
\begin{align}
\langle \tau^3 \rangle&= \int_N d\vN \int_x d\vx\, f_1 \tau_1^3 +
3\left(\langle \tau^2 \rangle -\langle \tau \rangle^2 \right)\langle \tau \rangle + \notag \\
&\phantom{\int}+ \int_N d^2\vN \int_x d^2\vx\, f_1 f_2 \xi_{12}(x_2 - x_1) \left( \tau_1^2 \tau_2
+ \tau_1 \tau_2^2 \right) \frac{3}{2} + \langle \tau \rangle^3 +
\notag \\
&\phantom{\int}+ \int_N d^3\vN \int_x d^3\vx\, f_1 \tau_1 f_2 \tau_2 f_3 \tau_3 \zeta_{123}(x_1,x_2,x_3).
\end{align}

The above moment calculations allow us to motivate how the series can be
re-summed, although see Appendix~\ref{app:poisson} for a proof that the resummation is exact at all orders in $\tau$ (as well as how to extend this calculation to higher order in the density).  Expressing the cumulants in terms of our expressions for the moments, inserting the cumulants into Eq.~\ref{eqn:mF}, and collecting terms yields
\begin{align}
\langle F \rangle &= \exp\left[ \int_N d\vN \int_x d\vx\, f_1 \left(-\tau_1
  + \frac{1}{2}\tau_1^2 - \frac{1}{3!}\tau_1^3 +\cdots \right) +
  \right.\notag \\
&\phantom{\int} + \frac{1}{2}\int_N d^2\vN \int_x d^2\vx\, f_1 f_2
  \xi_{12}(x_2 - x_1) \left(\tau_1 \tau_2 -
  \frac{1}{2}\left(\tau_1^2\tau_2 + \tau_1\tau_2^2\right) +
  \cdots\right) + \notag \\
&\left.\phantom{\int} + \frac{1}{3!}\int_N d^3\vN \int_x d^3\vx\, f_1 f_2 f_3 \zeta_{123}(x_1,x_2,x_3) \tau_1 \tau_2 \tau_3 +\cdots \right].
\label{eqn:mf1}
\end{align}
The parentheses in the first and second lines have forms that suggest they can be
re-summed into exponential functions of $\tau_i$.
Apart from the expansion in $\tau_i$, another expansion is in $\delta_L$, the
linear theory matter overdensity.  For ensuing calculations we cutoff at the
lowest nontrivial order in $\delta_L$ (i.e. quadratic order), as in simple halo
models.  In this case, only the two point correlation $\xi_{12}$ is nonzero, 
which we rewrite as $\xi_{12}^L$ to indicate that it is a biased tracer of the
linear matter field, and Eq.~\ref{eqn:mf1} becomes
\begin{align}
\langle F \rangle &= \exp\left[ \int_N d\vN \int_x d\vx\, f_1
  \left(e^{-\tau_1} - 1\right) +
  \right.\notag \\
&\left.\phantom{\int} + \frac{1}{2}\int_N d^2\vN \int_x d^2\vx\, f_1 f_2
  \xi^L_{12}(x_2 - x_1) \left(e^{-\tau_1} - 1\right) \left(e^{-\tau_2} -
  1\right)\right].
\label{eq:mean_F}
\end{align}
  Defining the effective optical depth $\tau_{\rm eff} =
-\ln{\langle F \rangle}$, our expression for the mean flux can be simplified to 
\begin{align}
\tau_{\rm eff} =  \overbrace{\int_N d\vN \int_x d\vx f_1 K_1(x_1)}^{\equiv \tau_{\rm eff}^P} 
\overbrace{- \frac{1}{2}\int_N d^2\vN \int_x d^2\vx\, f_1 f_2 \xi_{12}^L(x_1 + x_2) K_1(x_1) K_2(x_2)}^{\equiv \tau_{\rm eff}^C},
\label{eq:mF-simple}
\end{align}
where $K_i(x) \equiv 1 - e^{-\tau_a(x|N_i)}$.  Eq.~\ref{eq:mF-simple} is one of
the primary expressions used in this study.  The second clustering term,
$\tau_{\rm eff}^C$, is smaller by roughly the factor $\sim \tau_{\rm eff}^P
\xi(\sigma)$ relative to the first (Poissionian) term, $\tau_{\rm eff}^P$, and
$\xi(\sigma)$ denotes the $N$-averaged correlation function at the
characteristic linewidth.  Using the simulation described in the next section,
we find that the first term is larger by two orders of magnitude at the
redshifts we consider ($z=2.2$ and $z=3$, where $\tau_{\rm eff} \approx 0.16$
and $z=0.35$ respectively).

A similar calculation provides the two point correlation function. Starting from
the flux correlation function, 
\be \langle F(y) F(z) \rangle = \left \langle
e^{-\left[\tau(y) +\tau(z)\right]} \right \rangle.  
\ee
The two-point correlation of the flux field, $\langle F(y) F(z) \rangle $, is the same as the mean of the flux, $\langle F(y) \rangle$, with the replacement $\tau_1(y) \rightarrow \tau_1(y) + \tau_1(z)$ (c.f. eq.~\ref{eq:mean_F}). Thus, 
\begin{align}
\langle F(y) F(z) \rangle &= \exp\left[ \int_N d\vN \int_x d\vx\, f_1
  \left(e^{-\left[\tau_1(y) + \tau_1(z)\right]} - 1\right) +
  \right.\notag \\
&\left.\phantom{\int} + \frac{1}{2}\int_N d^2\vN \int_x d^2\vx\, f_1 f_2
  \xi_{12}^L(x_2 - x_1) \times \right. \notag \\
&\left.\phantom{\int} \times \left(e^{-\left[\tau_1(y) + \tau_1(z)\right]} -
  1\right) \left(e^{-\left[\tau_2(y) + \tau_2(z)\right]} -
  1\right) \right].
\label{eq:2point}
\end{align}

The statistic that is most commonly measured from the Ly$\alpha$ forest is the
correlation function of the flux overdensity, $\xi_F(y-z)$, defined as 
\be
\xi_F(y-z) \equiv \langle \delta_F(y) \delta_F(z) \rangle = \frac{\langle F(y)
  F(z)\rangle}{\langle F\rangle^2} -1,
\label{eq:xi_deltaF}
\ee 
where $\delta_F(x) = F(x)/\langle F \rangle - 1$.  Simplifying
Eq.~\ref{eq:2point} and decomposing it into an 1-absorber term and a
2-absorber term, using the notation 
\be 1 + \xi_F(r) = \exp\left[
\xi_\tau^{1a}(r) + \xi_\tau^{2a}(r) \right],
\label{eq:xiF_1a2a}
\ee
yields 
\begin{align}
\xi_\tau^{1a} &= \int_N d\vN \int_x d\vx\, f_1 K_1(x_1)
K_1(r-x_1), \label{eq:2point-1a}\\
\xi_\tau^{2a} &= \frac{1}{2} \int_N d^2\vN \int_x d^2\vx\,
f_1 f_2 \xi_{12}^L(r-x_1-x_2) \Big[ K_1(x_1) K_1(r-x_1)
  K_2(x_2) K_2(r-x_2) \Big. \notag \\
&\Big. - 2K_1(x_1)K_1(r-x_1)K_2(x_2) + 2K_1(x_1)K_2(x_2) \Big].
\label{eq:2point-2a}
\end{align}

As discussed in more detail shortly, the 1-absorber component will have a white
power spectrum on large scales, with a cutoff around the characteristic line
width \citep[see also][]{press93,zuo94}.  Eq.~\ref{eq:2point-2a} expressed the 2-absorber component into three
different terms in the integrand.  It turns out the largest term is the last one
($\propto K_1 K_2$).  This term convolves the line profiles $K$, with the
correlation function $\xi$, mirroring its 2-halo term analog in the
halo-model. The non-convolution (first and second) terms in $\xi_\tau^{2a}$ are
suppressed with respect to this term by a factor of $\sim K$ (or $\sim K^2$) in
the integrals, and on large scales contribute $20$\% and $10$\% of the 1D flux
power, respectively. Apart from their amplitude being suppressed, their
contribution also peaks on line-width scales. Additionally, due to their
differing signs, these two terms largely cancel on all scales, making their
total large-scale contribution $\sim 10$\%. We keep all terms in our numerical
calculations.

The previous formulation of the Absorber Model uses the line-of-sight profile of
absorbers and, hence, applies to line-of-sight correlations in the
forest. Extending it to include transverse correlations on the $\sim 0.1~$Mpc
size of absorbers would require modeling the transverse profile of $\tau$.
However, for widefield 3D Ly$\alpha$ surveys that are useful for large-scale
structure measurements, the transverse separation of sightlines is large enough
that modeling the transverse extent of individual absorbers is not relevant.
Instead, 3D surveys are sensitive to correlations between distinct absorbers
and, hence, measure our 2-absorber correlation function evaluated at some 3D
separation (such that the bias coefficients we derive in 1D apply in 3D).  The
next section computes these biases.

\subsection{Limits of the Absorber Model expressions}
\label{sec:simple_bias}
This section gives intuitive limits of the expressions we just derived.  It
helps to define the equivalent width of an absorption line with column $N$,
\be
{\mathrm{EW}}(N) \equiv \int dx K_1(x|N).
\ee 
The equivalent width is the effective size of an absorber.  Again ignoring the
subdominant clustering term in Eq.~\ref{eq:mF-simple} so $\tau_{\rm eff} = \tau_{\rm
  eff}^P$, the effective optical depth can be written as
\begin{equation}
 \tau_{\rm eff} = \int dN
f(N) \; {\mathrm{EW}}(N).
\label{eq:teff_poisson}
\end{equation}
Additionally, in the large-scale ($k\rightarrow0$) limit, the
1-absorber term can be thought of as the second moment of the equivalent width: 
\be
 P_{\tau}^{1a}(k\rightarrow 0) = \int dN
f(N) \;{\mathrm{EW}}(N)^2,
\label{eq:EW1a}
\ee where $P_{\tau}^{1a}$ is the Fourier transform of the 1-absorber
correlation function $\xi_{\tau}^{1a}$ (Eq.~\ref{eq:2point-1a}).  (See \citep{liske98} for a study of absorption correlations in this Poissonian limit.)  Similarly, the large-scale limit of the 2-absorber power
spectrum is 
\be 
P_{\tau}^{2a}(k \rightarrow 0) = P_L^{\rm 1D}(k) 
\left[\int dN \,f(N) b(N) {\mathrm{EW}}(N)\right]^2,
\label{eq:large_scale_2a}
\ee 
where $P_L^{\rm 1D}$ is the 1D power spectrum of $\delta_L$ (but we could also
have written 3D powers in above equation as the difference is just a projection
integral), we have used our result that the term with integrand over $2 K_1(x_1)
K_2(x_2)$ dominates Eq.~\ref{eq:2point-2a}, and we have also taken the linear
biasing relation $\xi^L_{12}(r) = b(N_1) b(N_2) \xi_L(r$), where $b(N)$ is the
bias of a line.  The latter linear-bias expansion for $\xi_{12}^L$
ignores redshift-space distortions, which we will incorporate soon.  It follows
(up to choice of sign) that the linear bias of the two absorber term, defined by
the relation $\delta_F \approx b_{\tau,\delta} \delta_L$ for $\delta_L \ll 1$,
is
\be
b_{F,\delta} = -\int dN f(N) b(N) {\mathrm{EW}}(N).
\label{eqn:bFd}
\ee 
A possible objection is that we took the limits of the terms in the
exponential and so really the above should be the optical depth bias,
remembering that $\xi_F = \exp[\xi^{1a}_\tau + \xi^{2a}_\tau] - 1$. Because
$\xi^{1a}_\tau $ equals zero outside the linewidth, on large-scales the
interaction terms between the 1-absorber and 2-absorber contribute a shot noise,
such that the flux density bias is the same as the optical depth density bias.
Appendix~\ref{sec:bias_beta} derives Eqn~\ref{eqn:bFd} using a different
approach that more formally shows that the flux density bias is the same as the
optical depth bias and also shows that the minus sign is correct.  The flux
density bias is identical to Eq.~\ref{eq:teff_poisson} for $\tau_{\rm eff}$
aside from the additional weighting by absorber bias, $b(N)$. Since $b(N)$ is
likely a weak function of $N$ (Appendix~\ref{sec:bias}), this suggests that similar column density systems dominate both the 2-absorber clustering signal and the Ly$\alpha$ absorption.  
 
The large-scale Ly$\alpha$ forest fluctuations are not sourced only be density
inhomogeneities but also redshift-space distortions owing to peculiar
velocities.  For discrete tracers of column $N_1$, the large-scale overdensity
is also perturbed by the gradient of the velocity field, such that on linear
scales $\delta(N_1) = b_1 \delta_L + \eta_L$, where $\eta_L \propto -\partial
v_L/\partial x$ \citep{1987MNRAS.227....1K}. Generalizing our large-scale
expansion of the flux field to

\be
\delta_F = b_{F,\delta} \delta_L + b_{F,\eta} \eta_L,
\ee
we can obtain $b_{F,\eta}$ from Eq.~\ref{eq:2point-2a} using similar logic to
how $b_{F,\delta}$ was derived, which yields

\be
b_{F,\eta} = - \int dN f(N) {\mathrm{EW}}(N) = -\tau_{\rm eff},\label{eqn:vbias}
\ee
again using that the optical depth bias and flux bias are equal. This second
expression identifies with our limit for the Ly$\alpha$ forest effective
optical depth (c.f. Eq.~\ref{eq:teff_poisson}) and is in accord at the
$\approx10\%$ level with the velocity gradient bias found in both observations
\citep{slosar11,bautista17} and simulations \citep{arinyo15}.  This result is not original, as
\cite{arinyo15} reasoned that $b_{F,\eta} = -\tau_{\rm eff}$ if only line
clustering is included.  There is an additional term in the velocity gradient
bias that our derivation ignored, and that was also reasoned to exist in
\cite{arinyo15}, which owes to the widths of lines depending on $\eta_L$ (as our
calculation only includes its effect on the clustering of absorbers). The full
expression is then

\be
b_{F,\eta} = - \int dN \, f(N)  {\mathrm{EW}}(N) \left[ 1 + \frac{\partial
    \ln{{\mathrm{EW}}(N)}}{\partial \eta_{\rm LS}}  \right] ,
    \label{eqn:bfeta}
\ee
where $\eta_{\rm LS}$ is the large-scale contribution to the velocity gradient.
See Appendix~\ref{app:poisson} for the full derivation.  The value of
$\partial{{\mathrm{EW}}}/{\partial \eta_{\rm LS}}$ depends on the model for the
linewidth.  We use a simple extension of our linewidth model in
Appendix~\ref{app:poisson} in which absorbers have size the Jeans length,
$\lambda_J$, and the velocity across the absorber is $\lambda_J H(z) (1 -
\eta_{\rm LS})$. This model yields the $\approx10-20\%$ correction (with $10\%$
at $z=2.2$ and $20\%$ at $z=3$) needed to bring our model into accordance with
the simulations and observations \citep{arinyo15}.

Seljak \cite{seljak12} derived the expression for velocity bias $b_{F,\eta} =
\langle F \ln{F} \rangle/\langle F \rangle$.  We find that at $z=2.2$ the
$b_{F,\eta}$ of \cite{seljak12} is 50\% smaller than ours.  The difference
between this bias and ours stems from different assumptions about the response
of the small scales to $\eta_L$. The result of \cite{seljak12} assumes simply
that the optical depth is affected by peculiar velocities via the mapping $\tau
\rightarrow \tau(1+\eta_L)$, whereas our approach implicitly assumed the
response of $\eta_L$ affects only large scales and not the small scales that are
shaped by Poisson statistics (which enter via the 1-absorber term).\footnote{The
  assumptions in our simplest $b_{F,\eta}$ derivation, which led to
  Eq.~\ref{eqn:vbias}, are the same as those that go into the Kaiser effect
  derivation for the galaxy clustering.} Our more sophisticated expression for
the bias (Eq.~\ref{eqn:bfeta}) improves upon our simplest model, including how
$\eta_L$ shapes the absorber linewidth.  \cite{cieplak15} found that the Seljak
formula is surprisingly accurate in the absence of thermal broadening (which
breaks the mapping $\tau \rightarrow \tau(1+\eta_L)$), likely because this
mapping is a reasonable approximation at densities relevant for the forest.
However, they found this expression undershoots by 30\% when thermal broadening
is included. Our biases account self consistently for thermal broadening.  (We
note that similar assumptions do not appear in the density bias derivation, as
there the density bias is treated more generally -- assuming that the flux is a
biased tracer of the matter.)

\section{Implementation of model}
\label{sec:implementation}
The Absorber Model is characterized by an
absorber column density distribution function $f(N_{\rm HI})$, a bias of the
absorbers $b(N_{\rm HI})$, and an optical depth profile $W(x|N_{\rm HI})$.  We use
numerical simulations and simple analytic models to characterize all three functions.
Numerical simulations are also essential for testing the Absorber Model.  This
section describes 1) the numerical simulations we use and 2) how we model
the absorber bias and optical depth profile.

\subsection{Numerical simulations and mocks}
\label{sec:mocks}

We create mock spectra based on high-resolution simulations.  The mocks are
constructed using the $z=2.2$ and $z=3.0$ snapshots of the reference model from
the Sherwood simulations suite \citep{bolton16}. This simulation was run with
$2\times 2048^3$ gas plus dark matter particles in a $40\Mpch$ box. The details
of the cosmological parameters and thermal history can be found in \citep{bolton16}.
For each simulation snapshot, the optical depth is calculated along 5000,
$40\Mpch$ sightlines through the box. The optical depth is further re-scaled, in
post-processing, to match the measured value of the mean flux from \citep{kim04}.

\begin{figure}[!ht]
  \centering
  \includegraphics[width=1.0\linewidth]{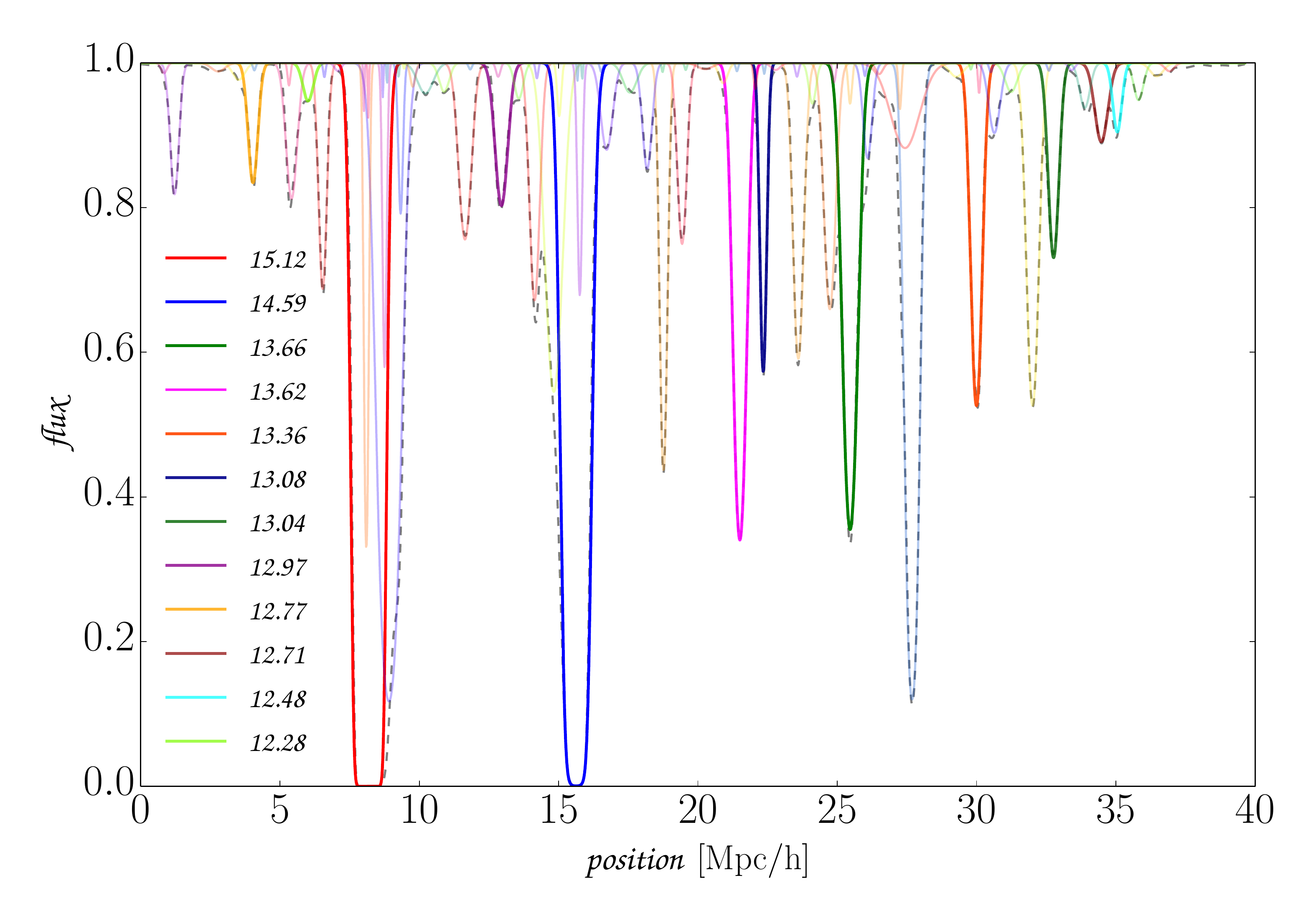}
  \caption{Showing absorber decomposition for a single skewer through our simulation
  box. The coloured curves show all the absorption lines fitted to the original
  simulation spectrum (shown in black dashed line). Twelve absorption profiles
  have been emphasized, spawning almost the full range of column densities
  probed. For those twelve, the colour coded legend displays their
  $\log_{10}(N_{\rm HI})$, using $\mathrm{cm^{-2}}$ units.}
  \label{fig:line_decomposition}    
\end{figure}

To decompose the simulated spectra into absorption lines, we have developed a
tool that decomposes the spectrum into lines with Gaussian optical depth
profiles, much like {\it VPFIT} \cite{vpfit}.  The algorithm starts with the
lines with the highest optical depths, fits and subtracts those, proceeds to the
remaining highest optical depth peaks, and so on.  To avoid overfitting, the
algorithm fits to a maximum of $80$ lines in each $40\Mpch$ skewer, and lines
with $N_{\rm HI} \leq 10^{11}\;\mathrm{cm^{-2}}$ are discarded from the
catalogue.  We find that the mean flux and power spectrum of this algorithm's
reconstructed flux field is essentially identical to the mean flux and power
spectrum of the simulation Ly$\alpha$ forest sightlines, indicating that the
algorithm's decomposition captures most of the absorption. This is also shown in
Fig.~\ref{fig:line_decomposition}, where all the lines fitted to a single skewer
are plotted (coloured curves). Black dashed line shows the result of the
original simulation \lya\ forest sightlines. In the figure, twelve typical
absorption systems were emphasized, corresponding to a range of column
densities, $N_{\rm HI} = 10^{12} - 10^{15}\;\mathrm{cm^{-2}}$. Sometimes the
method over-fits smaller systems in the simulation skewers but it is a minor
issue.

The spatial positions, column densities, and line widths from these fits are
used for $f(N_{\rm HI})$ and to test our $W(x|N_{\rm HI})$ model.  They are also
used to construct simplified mocks in addition to the full simulations.  In
particular, in addition to the simulation Ly$\alpha$ forest sightlines, this
study uses two simplified mock catalogues:
\begin{description}
\item [mocks:] These spectra use the line catalogue from fitting the simulation
  to reconstruct spectra, but substitutes the Absorber Model linewidth,
  $W(x|N_{\rm HI})$ (and discussed shortly) for the simulated linewidth.  This
  simplification is helpful for testing our model.
\item[random mocks:] Same as {\it mocks} except the positions of all lines have
  been randomly scrambled.  This eliminates the correlations between separate
  absorbers, making the 2-absorber term zero.
\end{description}

\begin{figure}[!ht]
  \centering
  \includegraphics[width=1.0\linewidth]{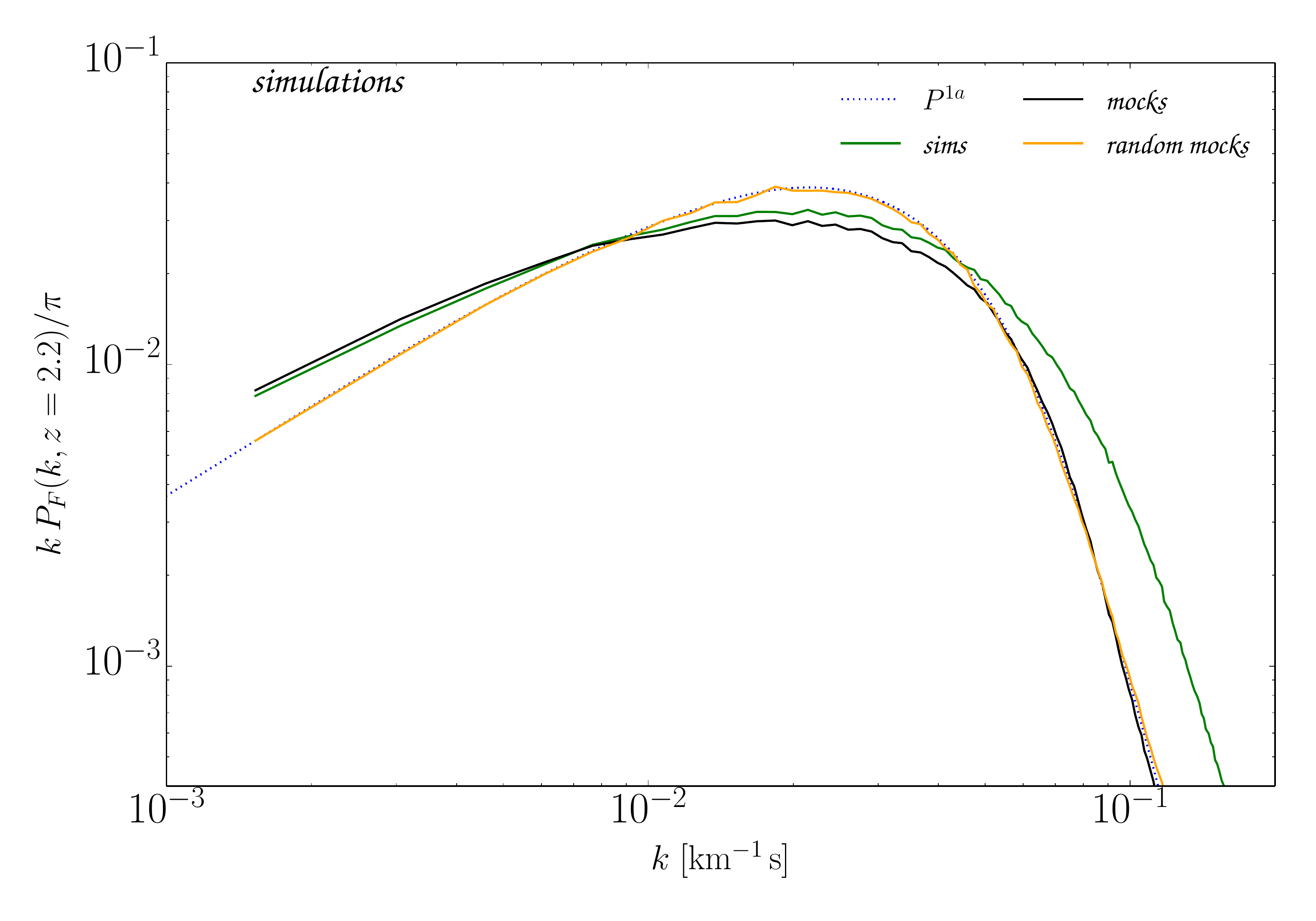}
  \caption{Comparison of the 1D flux power spectrum between spectra that are
    constructed from the simulation with different approximations.
    The green solid curve shows the power spectrum of the simulation spectra, and the black solid curve
    shows the power spectrum of the {\it mocks} spectra.  The {\it mocks} used the fitted positions and columns in
    the simulations but substitute our simple linewidth model. The orange solid curve
    shows the power spectrum of the {\it random mocks}, which are the same as the {\it
      mocks} except that the positions of the absorption features have been
    randomly scrambled to eliminate correlations.  See Section~\ref{sec:mocks} for
    additional description. The blue dotted curve is the
    model 1-absorber contribution.  Up to sample variance
    in the simulation, this analytic 1-absorber calculation should be identical to the {\it random
      mocks} prediction.  }
  \label{fig:pk_sims}    
\end{figure}

The differences between the full simulations (denoted {\it sims} in plots) and
these simplified mocks are instructive for understanding what drives
correlations.  At our fiducial redshift of $z=2.2$, the full simulation spectra
have $\tau_{\rm eff}= 0.160$ and the {\it mocks} spectra (which use our line
model) have $\tau_{\rm eff}= 0.169$.  As alluded to earlier, we find that
randomizing the positions of absorbers in the simulations results in sub-percent
differences in $\tau_{\rm eff}$ (at $z=2.2$).  Additionally, Fig.~\ref{fig:pk_sims} shows the
changes in the power spectrum of the normalized flux overdensity between the
full simulation ({\it sims}) and the same, but substituting our line-width model
({\it mocks}). Most of the differences in the power spectrum between the {\it
  sims} and {\it mocks} lies in the thermal cut-off at high wavenumbers.
We believe this occurs because the {\it sims} capture the scatter in line widths at fixed $N_{\rm HI}$, while
the {\it mocks} assume a one-to-one correspondence between the two given
approximately by the mean relation.  However, the large-scale behavior of the
power spectrum is only altered at the few percent level between {\it sims} and
{\it mocks}, suggesting that the line width model is less important there.

The yellow solid curve in Fig.~\ref{fig:pk_sims} is the flux power spectrum obtained from the
 {\it random mocks}.  The {\it random mocks} power spectrum is more discrepant with the
 {\it sims} and {\it mocks} power spectra.  This indicates that the correlations between
 distinct absorbers have some effect on shaping the Ly$\alpha$ forest power
 spectrum.  We note that, up to sample variance, the random mocks power spectrum
 should be identical to the 1-absorber term in our model.  This result is confirmed in Fig.~\ref{fig:pk_sims} by comparing the yellow solid {\it random mocks} curve with the blue dotted
 1-absorber curve (which is calculated from solving our Eq.~\ref{eq:2point-1a} and the same line width model).

\subsection{Absorber properties in models}
\label{ss:modelprops}

The simulation's line catalogue plus simple analytic models are used to
construct the inputs for the Absorber Model.  Here we describe each input:

\begin{description}
\item[column density distribution:] The column density distribution function,
  $f(N_{\rm HI})$, is tabulated from our simulation's absorber catalogue.  Note
  that the simulations largely miss $N_{\rm HI} \gtrsim
  10^{17}\;\mathrm{cm^{-2}}$ because they do not have self-shielding; this does
  not affect our comparison between the simulations and the Absorber Model, but
  does result in the calculations in this paper missing the not-insignificant
  contamination from high-$N_{\rm HI}$ absorbers \citep{viel04DLA,font12b, mcdonald05b,
    mcquinnwhite, 2017arXiv170608532R}.

\item[line profile:] We assume that $W(x|N_{\rm HI})$ is Gaussian with standard
  deviation $\sigma_a(N_{\rm HI})$.  To model $\sigma_a$, we use a model based
  on that presented in \citep{garzilli15b}.  This model uses the correspondence
  between the column density $N_{\rm HI}$, gas over-density $\Delta_b$, and gas
  temperature $T$ found in simulations \citep{mcquinn11,
    altay11} and that applies to the extent that the size of
  absorbers is set by the Jeans length \citep{schaye01} and that the temperature
  follows a $T-\Delta$ relation \citep{hui97, 2016MNRAS.456...47M}.
  Additionally, this model makes the ansatz that the velocity broadening is set
  by applying Hubble's law across the Jeans length extent of the absorbers.
  This velocity broadening plus thermal Doppler broadening are added in
  quadrature to set the linewidth.  (The velocity broadening is really determined by
  the nonlinear velocity field; however, at the low densities where Hubble
  broadening likely dominates over thermal Doppler, this Hubble-broadening approximation is more
  relevant.)  This model captures the mean of the distribution of linewidths in
  simulations \citep{garzilli15b}.  The model ignores the wide dispersion in
  linewidths at fixed column, although this dispersion appears to primarily
  affect the 1D power spectrum near the thermal cutoff (compare {\it sims} and
  {\it mocks} curves in Fig.~\ref{fig:pk_sims}; the {\it mocks} uses this line
  model).  See Appendix~\ref{sec:thermal_model} for more details on this line
  model.  Finally, we assume that $T = T_0 \Delta_b^{\gamma-1}$, with $T_0 =
  12,300\;\mathrm{K}$ and $\gamma = 1.59$ to match our simulations at $z=2.2$.

\item[linear bias]: Our model for the absorber bias is based on a one-to-one
  matching of a Lagrangian region with linear overdensity $\delta_L$ with a
  Lagrangian region that results in a column of $N_{\rm HI}$ and Jeans-length
  physical size $\lambda_J(N_{\rm HI})$, with this matching ordered in increasing
  $\delta_L$ and $N_{\rm HI}$.  However, in reality not all gas is accounted for
  in our fits to the column density distribution.  Systems that contribute weak
  absorption are missed because of incompleteness of the line fitting as well as
  because the absorption-line description breaks down for extremely diffuse gas.
  Systems in dense regions are missed, for one, because $\Delta_b \sim100$ gas
  tends to shock heat at virialization and, hence, not host significant neutral
  hydrogen.  Because the missing gas tends to be at the lowest and highest
  densities that have the large absolute bias factors, we find that accounting
  for these missing contributions is important in order for the Absorber Model to
  match the large-scale power spectrum.

Therefore, we define parameters $\epsilon_<$ and $\epsilon_>$ that account for
the breakdown of this one-to-one mapping at low and high columns (see
Appendix~\ref{sec:bias} for more details).  We find that setting $\epsilon_< =
10^{-2}$ eliminates the divergence to large negative biases that appears for
much smaller $\epsilon_<$, and that the results are stationary when changing
$\epsilon_<$ by a factor of several around this value.  Thus, in our bias models
we fix $\epsilon_< = 10^{-2}$ and we adjust $\epsilon_> $.  The fraction of
highly biased material that is not accounted for by our column density
distribution is given by $\epsilon_>/(1.5+\epsilon_>)$ at $z=2.2$
(Appendix~\ref{sec:bias}).  Simulations show that $20\%$ of the gas in the
Universe shock heats to $T>10^5$K \citep{cen06}, collisionally
ionizing much of the neutral hydrogen, and relatedly, at $z=2.2$, $20\%$ of the
gas is in $>10^{10} M_\odot$ dark matter halos that are able to pull in
gas and form galaxies.  Thus, our rough expectation is that $20\%$ of dense gas
is likely ``missing'' in our column density distribution, meaning
$\epsilon_>/(1.5+\epsilon_>) \sim 0.2$ or equivalently $\epsilon_> \sim 0.5$.
We find in Section~\ref{ss:bias} that $\epsilon_> = 0.5-1$ is needed to match the
large-scale power in the simulations, with the exact value depending on the
absorber exclusion model.  Figure~\ref{fig:bias_epsilon} in
Appendix~\ref{sec:bias} shows that for $\epsilon_> = 0.9$ the Eulerian absorber
bias ranges from $b(N_{\rm HI}) \approx -0.5$ at $N_{\rm HI} = 10^{11-12}$cm$^{-2}$
to $b(N_{\rm HI}) \approx 0.8$ for $N_{\rm HI} = 10^{14.5-17}$cm$^{-2}$ (with
the latter tending to $b(N_{\rm HI}) \approx 1.1$ if we instead take $\epsilon_>
= 0.4$).

\end{description}

In this paper we also assume that the linear gas density is a windowed version
of the linear total matter density due to pressure smoothing, such that $P_b =
P_L e^{-k^2/k_F^2}$ where $P_b$ and $P_L$ are the linear gas power spectrum and
linear total matter power spectrum, respectively, and $k_F = 2\pi/\lambda_F$ is
the filtering scale which we take to be $k_F^{-1}=0.041\;\cMpch$
(\citep{gnedin98,kulkarni15}; see the appendices for details).
 However, the Absorber Model flux power spectrum is insensitive to this
 $k_F$-windowing because the small-scale power is dominated by the 1-absorber
 term at the redshifts we have analyzed.

These choices complete the inputs needed to compute correlations in our model
using Eqs.~\ref{eq:xiF_1a2a}-\ref{eq:2point-2a}.  In what follows, we calculate
the power spectrum in the model and compare with the power spectrum from the
simulation mocks.

\section{The Ly$\alpha$ forest power spectrum}
\label{sec:main}

\begin{figure}[!ht]
  \centering
  \includegraphics[width=1.0\linewidth]{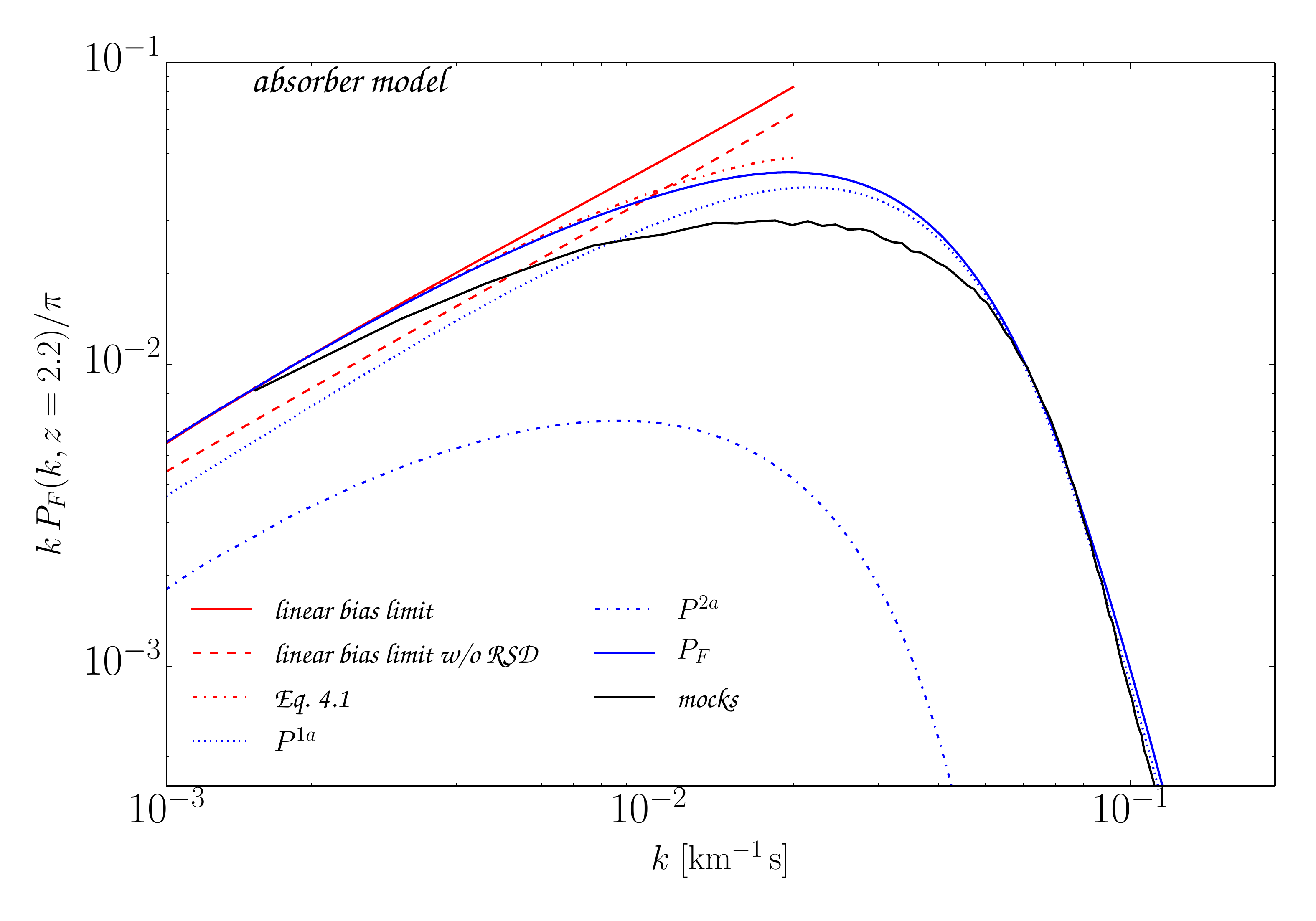}
  \caption{ Flux power spectra as calculated in the simulation {\it mocks} (black
    solid curve) and in the Absorber Model (blue curves).  The Absorber Model is
    decomposed into contributions from the 1-absorber term (blue dotted curve),
    the 2-absorber term (blue dot-dashed curve), and their total (blue solid
    curve).  The red solid line shows the large-scale limit of the Absorber Model using
    Eq.~\ref{eq:linear_bias}, the red dashed line shows the same but without redshift-space distortions, and the dot dashed curve shows this calculation keeping the scale dependence of $N_p^{-1}$.}
\label{fig:pk_model}
\end{figure}

Now that we have all the ingredients needed to compute the correlations in our
model, we focus on the line of sight (or 1D) flux power spectrum.
Fig.~\ref{fig:pk_model} show the different components that contribute to the
absorber power spectrum.  Respectively, the 1-absorber, 2-absorber, and full
power spectrum were calculated by taking the Fourier transform of
Eq.~\ref{eq:2point-1a}, Eq.~\ref{eq:2point-2a} and Eq.~\ref{eq:xiF_1a2a}, with
the models for $f(N_{\rm HI})$, $W(x | N_{\rm HI})$, and $b(N_{\rm HI})$
specified in the previous section.  The overall shape of the Absorber Model
power spectrum agrees to $\sim 20\%$ with the mock spectra, with the best
agreement on the small scales that are determined by the 1-absorber term and the
worst at intermediate scales.  Since we are comparing to the {\it mocks}, and not the
original simulations, the only two effects that can change the behavior of the
flux power at those scales is non-linear clustering and the related positional
`exclusion' of absorption lines (discussed shortly).

Interestingly, the 1-absorber term is at least three times larger than the
2-absorber term, with the difference increasing with wavenumber.  On scales much
larger than the linewidth, the 1-absorber power spectrum is white.  This has a
couple interesting implications.  First, mode counting arguments for the
constraining power of the Ly$\alpha$ forest have missed this source of noise, and previous attempts to use the 1D power to measure the bias of the
transmitted flux have overestimated the bias because of the 1-absorber term
\citep{mcquinnwhite, mcdonald03}.

Let us understand the large-scale limit of the power spectrum. At large-scales
$\xi_F = \exp[ \xi_\tau^{1a} + \xi_\tau^{2a}]-1$ can be expanded to yield $\xi_F
\approx \exp[ \xi_\tau^{1a}](1+  \xi_\tau^{2a})-1$, or taking the Fourier
transform:

\be
P_F(k) = P_{\tau}^{2a}(k) + P_F^{1a}(k) + P_{\tau}^{2a}(k)
\star P_F^{1a}(k),
\label{eq:large_scale_1}
\ee
where $P_F^{1a}(k)$ is the power spectrum of $\xi_F^{1a} = \exp[ \xi_\tau^{1a}]$
(Eq.~\ref{eq:2point-1a}), which asymptotes to a constant at low $k$, and
$P_{\tau}^{2a}$ is the Fourier transform of $\xi_\tau^{1a}$
(Eq.~\ref{eq:2point-2a}). The last term in Eq.~\ref{eq:large_scale_1} is just a
few percent of the total power at all wavenumbers, and, if it is dropped, we can
rewrite Eq.~\ref{eq:large_scale_1} as
\be
P_F(k) \approx \int_k^\infty\, dq\, q\, P_m^{3D}(q) \left(b_{F,\delta}
+ f b_{F,\eta} \mu^2 \right)^2 + N_P^{-1},
\label{eq:linear_bias}
\ee
where $\mu = k/q$ is the angle between the line of sight and the Fourier vector
$\vec{q}$, and the linear bias factors for the density ($b_{F,\delta}$) and
velocity gradient ($b_{F,\eta}$) are given by the relations in
Sec.~\ref{sec:simple_bias}. The $k$-independent Poissonian term, $N_P^{-1} \approx
P_F^{1a}(k\rightarrow 0)$, is dominated by the Poisson 1-absorber term.\footnote{Apart
from the Poisson 1-absorber term, both non-convolution terms from 2-absorbers
act as additional source of shot-noise on large scales, modulated by weighted
integrals of the matter power spectrum.}  In the $\tau_{\rm eff}\ll1$ limit, the
shot-noise term is related to the effective optical depth through
Eq.~\ref{eq:EW1a}. The solid red line in
Fig.~\ref{fig:pk_model} is the evaluation of Eq.~\ref{eq:linear_bias} for our model, where all the non-convolution terms of
2-absorber power (which have a white power as $k\rightarrow0$) were kept in $N_P^{-1}$.  If we keep the
scale dependence of $N_P^{-1}$, the result (the dot-dashed red curve) deviates significantly from the red solid line only at $k\gtrsim0.01\,\skm$ and is only 12\% larger at $k=0.02\,\skm$ than the full model flux power (full blue line in Fig.~\ref{fig:pk_model}). The approximation of
Eq.~\ref{eq:linear_bias} but with $b_{F,\eta} = 0$ is
shown as dashed red line in Fig.~\ref{fig:pk_model}, showing that redshift-space distortions shape the power at the 20\% level (contributing at ${\cal O}(0.4)$ level to the 2-absorber term in the Reference Model). 

\begin{figure}[!ht]
  \centering
  \includegraphics[width=1.0\linewidth]{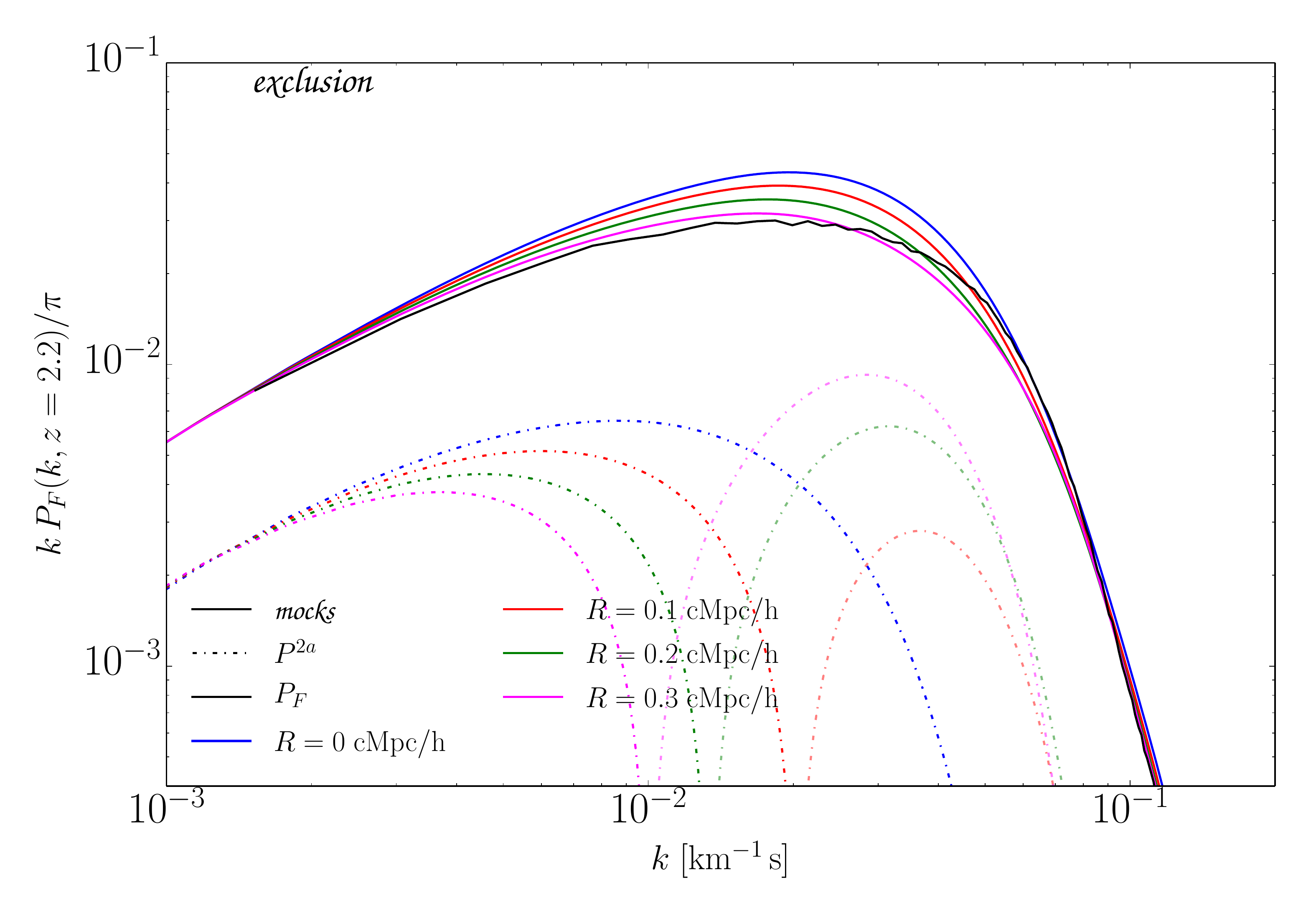}
  \caption{ The effect of absorber exclusion on the flux power spectrum. The
    black solid curve corresponds to the results of the simulation {\it mocks}, while different
    colored curves correspond to setting $\xi_{12}(x)=-1$ at $x<R$, with $R = 0$ (in blue) being the Reference
    model. The colored solid curves show the total flux power in the absorber
    model, while the dot-dashed curves with corresponding color shows the
    contribution from the 2-absorber term only.  The translucent curves indicate
    regions of negative 2-absorber power. For all models, the bias has been
    tuned to match the large-scale limit of the {\it mocks}.
  }
\label{fig:pk_exclusion}
\end{figure}

\subsection{Absorber exclusion}  
Absorbers' positions can be anti-correlated on small scales owing to absorbers having to occupy different real space positions -- absorber
exclusion.
We show that absorber exclusion is likely the reason why our Reference Model,
which does not include exclusion, overpredicts the 1D power spectrum at
intermediate wavenumbers.

We illustrate exclusion with a simple model in which we remap the absorber
correlation function in the 2-absorber term (Eq.~\ref{eq:2point-2a}) to

\be
\xi_{12}^L(x) \rightarrow 
\begin{cases} 
-1 & \mbox{if } 0 < x < R; \\
b_1 b_2 \xi_L(x) & \mbox{if } x \ge R,
\end{cases}
\ee
where $\xi_L$ is the usual linear matter correlation function.  While we have
not included redshift-space distortions in the above, really for $x \ge R$ we
include these terms but set them to zero at $x<R$ (so that our exclusion
happens in redshift space).
Figure~\ref{fig:pk_exclusion} shows the effect of exclusion on the 1D power
spectrum.  The different curves vary $R$ by up to a few times the
Jeans length at $z=2.2$ ($\lambda_J \sim 0.05 \,\cMpch$; Eq.~\ref{eq:lJ}).  The Jeans length
is the characteristic size of absorbers. (Analogously, to explain
the sub-Poissonian shot noise found in halo clustering, halo exclusion is found
to become important below a few times the virial radius \citep{baldauf13}.)  The $R = 0\,\cMpch$ model is our Reference Model, which
does not include exclusion. While
$R=0.3\;\cMpch$ works well on large scales, matching the {\it mocks} on small scales requires a smaller value
of $R$ (see Fig.~\ref{fig:pk_exclusion}), a trend likely reflecting the simplicity of the exclusion model used.

Additionally, the absorber exclusion suppresses the large-scale power in
the model by reducing the shot noise. The actual effect is coming from a
modified 1D matter power spectrum. For a simple model of line exclusion
considered in this paper, an analytic relation can be obtained such that
\be
P_{12}(k;R) = P_{12}(k;R=0) - 2R\, W_R(k) - 2R\, \left[ P_{12} \star W_R
  \right](k),
\label{eq:line_exclusion}
\ee
where $P_{12}(k)$ is the Fourier transform of $\xi_{12}^L$ and $W_R(k) =
\sin(kR)/(kR)$ is a 1D Fourier transform of a top-hat function and $R$ is the
exclusion scale. In the large-scale limit, $k \rightarrow 0$, the correction due
to absorber exclusion in the flux power spectrum can be written as
\be
P_F^{2a}(k;R) = P_F^{2a}(k;R=0) - \lambda \,R,
\label{eq:simple_exclusion}
\ee
and $\lambda=0.09$ in our model. The convolution term from exclusion in Eq.~\ref{eq:line_exclusion} adds less than 4\% of the total value of
$\lambda$. 
 However, the overall effect of exclusion is quite large:  in Eq.~\ref{eq:simple_exclusion}, $\lambda\,R$ is 75\% of $P_F^{2a}(R=0)$ for $R=0.3\;\cMpch$.\footnote{The effect of the
  line exclusion can also be observed in the effective optical depth, through
  the contribution of the clustering term $\tau_{\rm eff}^C$. The result to
  which we have already alluded to before, is that at lower redshifts when the
  total effective optical depth is small enough, this contribution is less than
  a percent. However, evaluating $\tau_{\rm eff}^C$ with the full linear matter
  correlation function (without line exclusion), results in large changes in the
  optical depth, even at low redshifts. Even as in the flux power spectrum, in
  Fig.~\ref{fig:pk_exclusion}, the preferred values of $R$ to match the
  effective optical depth of the model with the {\it mocks}, are of the order of $R =
  0.2 - 0.3\,\cMpch$, which is a few times the typical scale of the absorbers.}

Fig.~\ref{fig:pk_exclusion}, however, does not show the large-scale suppression of power from exclusion.  Since this effect is perfectly degenerate with the choice of large-scale
absorber bias, $b_{F,\delta}$, and we take that effect out of the calculation
by changing the biasing scheme. To match the large-scale limit, the adjusted bias values for $R=0.1, ~0.2$ and
$0.3\;\cMpch$ are $b_{F,\delta}=-0.09,\, -0.1$ and $-0.11$ respectively. This should be compared
to the value of bias in the Reference Model, $b_{F,\delta}=-0.08$.

\subsection{Absorber bias}
\label{ss:bias}
\begin{figure}[!ht]
  \centering
  \includegraphics[width=1.0\linewidth]{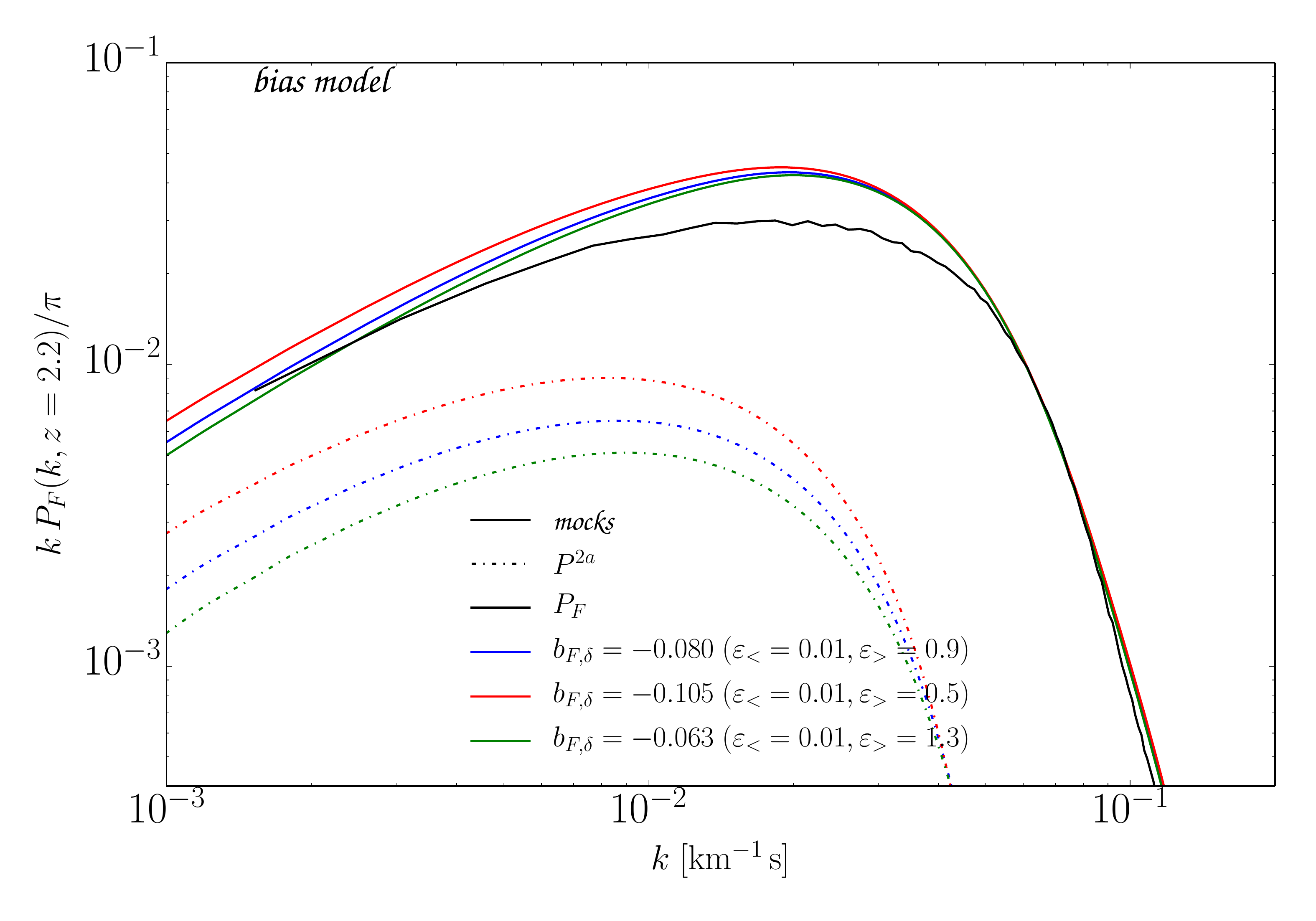}
  \caption{The effect of varying the parameters of our bias model on the 1D flux
    power spectrum. Different colored curves have varied our simple bias model's two free parameters, $(\varepsilon_>,\varepsilon_<)$, with solid indicating the total power and the dot-dashed indicating the 2-absorber contribution. The
    full black curve corresponds to the results of the simulation {\it mocks}.}
\label{fig:pk_bias}
\end{figure}

In our simple model of absorber bias we can vary two parameters, $\varepsilon_>$
and $\varepsilon_<$, to match the large-scale bias in the {\it mocks} spectra
(see Appendix~\ref{sec:bias}). Of these, $\varepsilon_>$ is the most important,
representing the amount of highly biased material that does not contribute to
the forest absorption (because of, e.g., shock heating).  In our Reference
model, these parameters were chosen to be $(\varepsilon_<,~\varepsilon_>)=
(0.01,0.9)$ such that the large-scale 1D flux power spectrum of the model agrees
with the results from the simulation {\it mocks} at $z=2.2$. Here we investigate
the impact of varying $\varepsilon_>$.  We show $\varepsilon_> = \{0.9, ~0.7,~
1.3\}$ in Fig.~\ref{fig:pk_bias}, which respectively result in $b_{F,\delta} =
-\{0.080,~0.105,~0.063\}$. Thus, varying $\varepsilon_>$ over this range results
in a small change in $b_{F,\delta}$ and, as a result, small changes in the
large-scale power.  The model with $\varepsilon_> = 0.9$ is most consistent with
the {\it mocks} power.  This is consistent with the $\varepsilon_> = 0.5-1$ we
motivated in Section~\ref{ss:modelprops} (from considering gas `lost' to shock
heating and galaxy formation), and we note that the models with exclusion that
match the bias require $\varepsilon_> = 0.71, ~0.57,$ and $0.46$ for $R = 0.1,
~0.2$ and $0.3\;\cMpch$ respectively.  Thus while a certain degree of tuning was
performed on the bias to match the large scale limit of the {\it mocks} power,
the chosen parameter values lie well within the reasonable assumption of the
simple bias model used. We suspect improved models for the absorber bias and
absorber exclusion can be developed, models for which less tuning is required.

Since we find a best fit flux bias of $b_{F,\delta} = -0.08$, this translates
into the effective bias of the absorption lines, $-b_{F,\delta}/\tau_{\rm eff}$
(c.f.~Eq.~\ref{eqn:bFd}), of approximately $0.5$, using $\tau_{\rm eff} = 0.16$.
This is approximately the model bias of $N_{\rm HI} \sim 10^{14}~$cm$^{-2}$
systems that Section~\ref{sec:pdf_NHI} shows dominate the 2-absorber
correlations.

\subsection{Redshift evolution}
\label{sec:redshift_evo}

The amplitude of the 1D Ly$\alpha$ forest power spectrum increases significantly
with redshift, and constraints on cosmological parameters derive a lot of their
sensitivity from having measurements at multiple redshifts.  Thus, it is
interesting to consider our model at a higher redshift to determine whether it
captures the expected trends.  We again compare the Absorber Model with the {\it
  mocks} spectra (which share the same linewidth model), but these spectra are
now constructed using the $z=3.0$ simulation output. We also use the same bias
parameters $(\varepsilon_>,\varepsilon_<)$ as the $z=2.2$ Reference Model.

\begin{figure}[!h]
  \centering
  \includegraphics[width=1.0\linewidth]{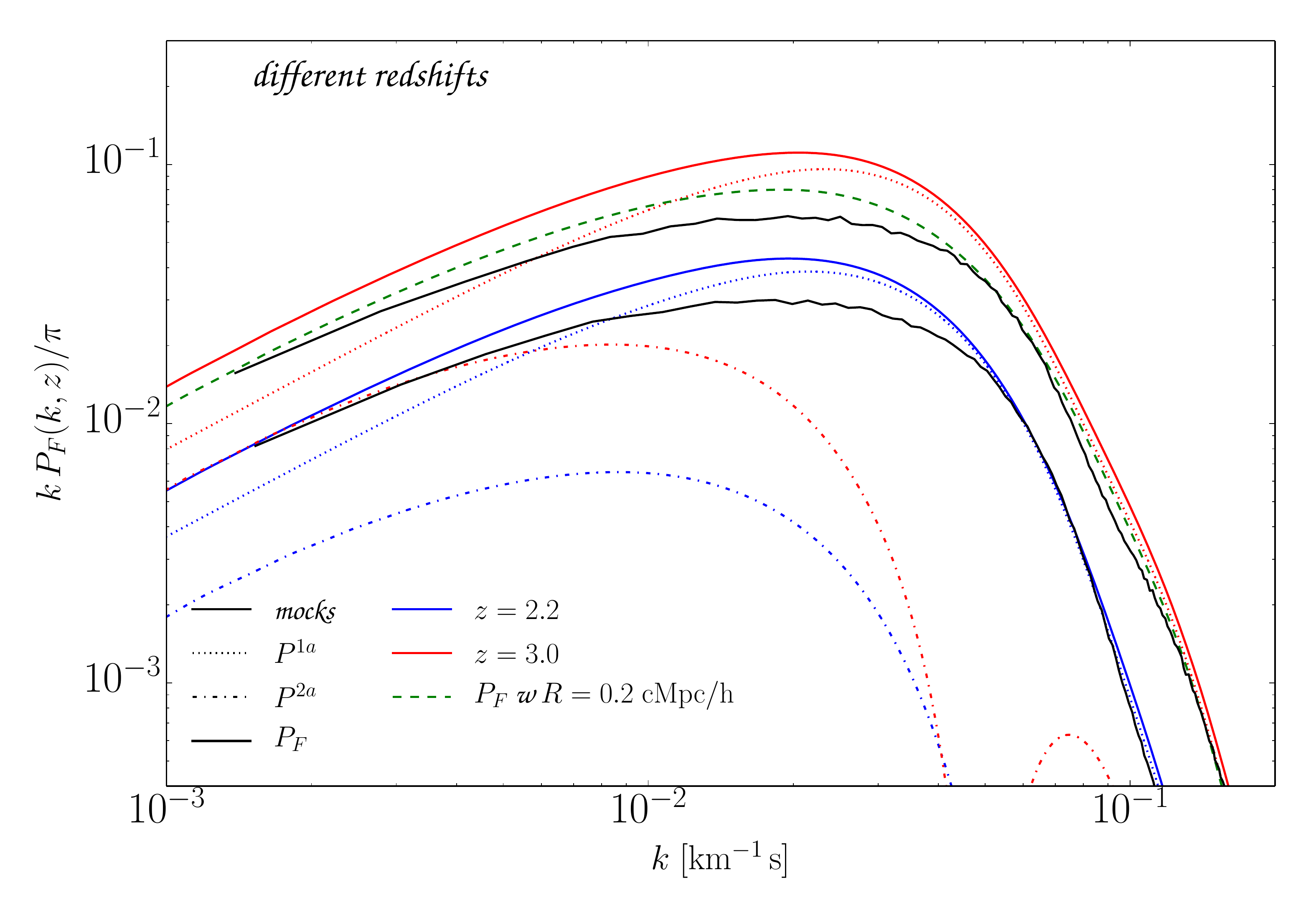}
  \caption{ Comparison of 1D flux power spectrum at $z=2.2$ (in blue) and
    $z=3.0$ (in red). The blue and red solid curves show the total flux power
    spectrum in the Absorber Model, while the dotted and dot-dashed lines show
    contribution from 1-absorber and 2-absorber terms, respectively. The black
    solid curves shows the results of the simulation {\it mocks}, with the
    $z=2.2$ case below $z=3.0$. The green dashed curve shows at $z=3$ model with
    line exclusion below $R = 0.2\;\cMpch$.}
\label{fig:pk_z}
\end{figure}

Fig.~\ref{fig:pk_z} compares the flux power spectra of the Absorber Model and
{\it mocks} for both $z=2.2$ (lower curves) and $z=3.0$ (upper curves).  By and
large, the $z=3.0$ curves all are shifted up by a comparable factor in amplitude
compared to the $z=2.2$.  Thus, our previous conclusions apply: The 1-absorber
term is similarly dominant.  The effect of clustering is similar, albeit
somewhat larger at $z=3.0$.  The ratios of the total power, 1-absorber power,
and 2-absorber power at $k=10^{-3}\;\skm$, are $1$~:~$0.66$~:~$0.32$ at $z=2.2$
and $1$~:~$0.57$~:~$0.40$ at $z=3$. (The convolution term in
Eq.~\ref{eq:large_scale_1} is 1\% and 2\% of the large-scale power at $z=2.2$
and $3.0$, respectively.) The Absorber Model is slightly more discrepant with the {\it
  mocks} at $z=3.0$, overpredicting the large-scales power by $\approx15$\%.
While our bias model was tuned to match large-scale properties at $z=2.2$, we
have not tuned parameters for $z=3.0$. Indeed, we find that decreasing the
value of $b_{F,\delta}$ in the model by $30$\% matches the large-scale power in
the {\it mocks} at $z=3.0$.

Both observations \citep{mcdonald05,mcdonald06} and simulations
\citep{mcdonald03,arinyo15} suggest redshift evolution to be power-law with
redshift, with the power-law index of slightly less than $3$ ($b_{F,\delta}
\propto (1+z)^{2.9}$). The results of the Absorber Model suggest a redshift
evolution closer to a power-law index of $2$ ($b_{F,\delta} \propto (1+z)^2$),
with the caveat that this scaling is likely affected by the thermal history.
Using the $30$\% lower value of $b_{F,\delta}$ suggested by the large-scale
limit of the $z=3.0$ {\it mocks} power would further increase the discrepancy
with the observed power-law index. Increasing the amount of absorber exclusion
with redshift could help reconcile this discrepancy.  (Fixing the large-scale
power results in $\approx 30\%$ larger $b_{F,\delta}$ to compensate for
exclusion with $R=0.2\;\cMpch$. The green solid curve in Fig.~\ref{fig:pk_z}
shows our $R=0.2\;\cMpch$ toy exclusion model but with the same $b_{F,\delta}$.)
Larger model biases could also be created by decreasing $\varepsilon_>$, which
is theoretically motivated as only half the gas has shocked heated or collapsed
into galaxy-sized halos at $z=3$ compared to $z=2.2$ (\cite{cen06}; suggesting a
similar reduction in $\varepsilon_>$). Thus, the discrepancy at large scales at
$z=3.0$ falls within the plausible range of input parameters to the Absorber
Model.


\subsection{Column densities}
\label{sec:pdf_NHI}

\begin{figure}[!h]
  \centering
  \includegraphics[width=1.0\linewidth]{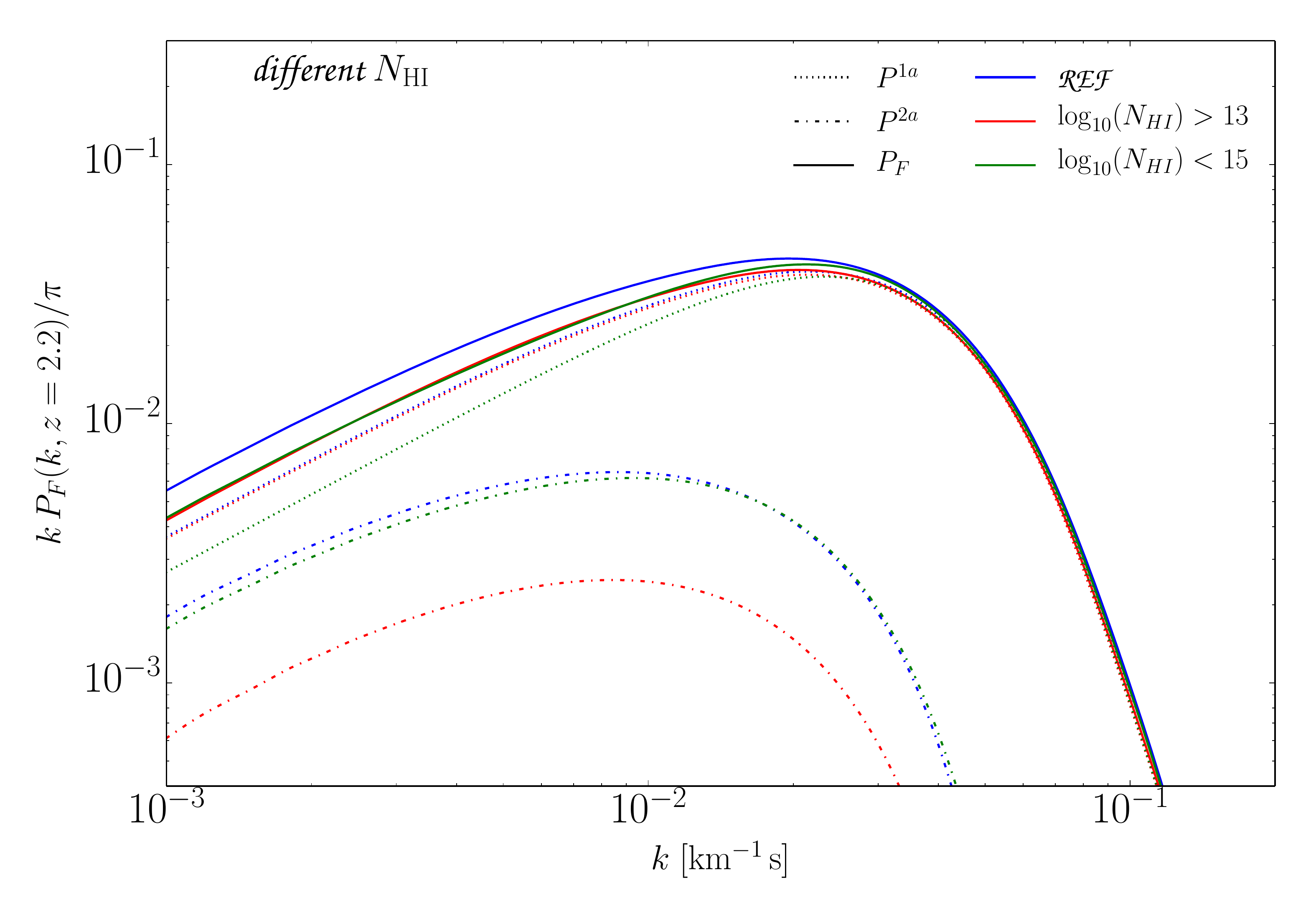}
  \caption{ Absorber model power spectrum for different column density cuts. The
    solid curves show the total flux power, the dot-dashed curves show the
    2-absorber term, and the
    dotted curves show the 1-absorber term. The Reference Model is in blue,
    which includes all columns,
    $11 \le \log_{10}(N_{\rm HI}) \le 17$. The flux power excluding low column
    densities is shown in red, and the flux power excluding high column
    densities in green.}
\label{fig:pk_Ncuts}
\end{figure}

The Absorber Model describes the \lya\ forest by discretizing it into systems of
different columns.  This discretization allows us to understand which
systems are most important in shaping the correlations.  Fig.~\ref{fig:pk_Ncuts}
shows the predictions of the Absorber Model when making different cuts on column
density. The Reference Model (the blue curve) uses the range between $11 \le
\log_{10}(N_{\rm HI}) \le 17$, using cm$^{-2}$ units. When excluding the low
column densities from the calculation with $\log_{10}(N_{\rm HI}) < 13$ (solid
red curve), the flux power spectrum is reduced by $13\%$ on large
scales. However, the small-scale power is nearly unchanged. Excluding low
columns also has little effect on 1-absorber term, whereas the 2-absorber term
is reduced by a factor of several when excluding the columns below
$\log_{10}(N_{\rm HI}) = 13$ (red lines). Excluding the high column densities
with $\log_{10}(N_{\rm HI}) > 15$ also mostly affects large scales.  However,
for this case, the 1-absorber term is most changed, being reduced by $17\%$.

\begin{figure}[!h]
  \centering
  \includegraphics[width=1.0\linewidth]{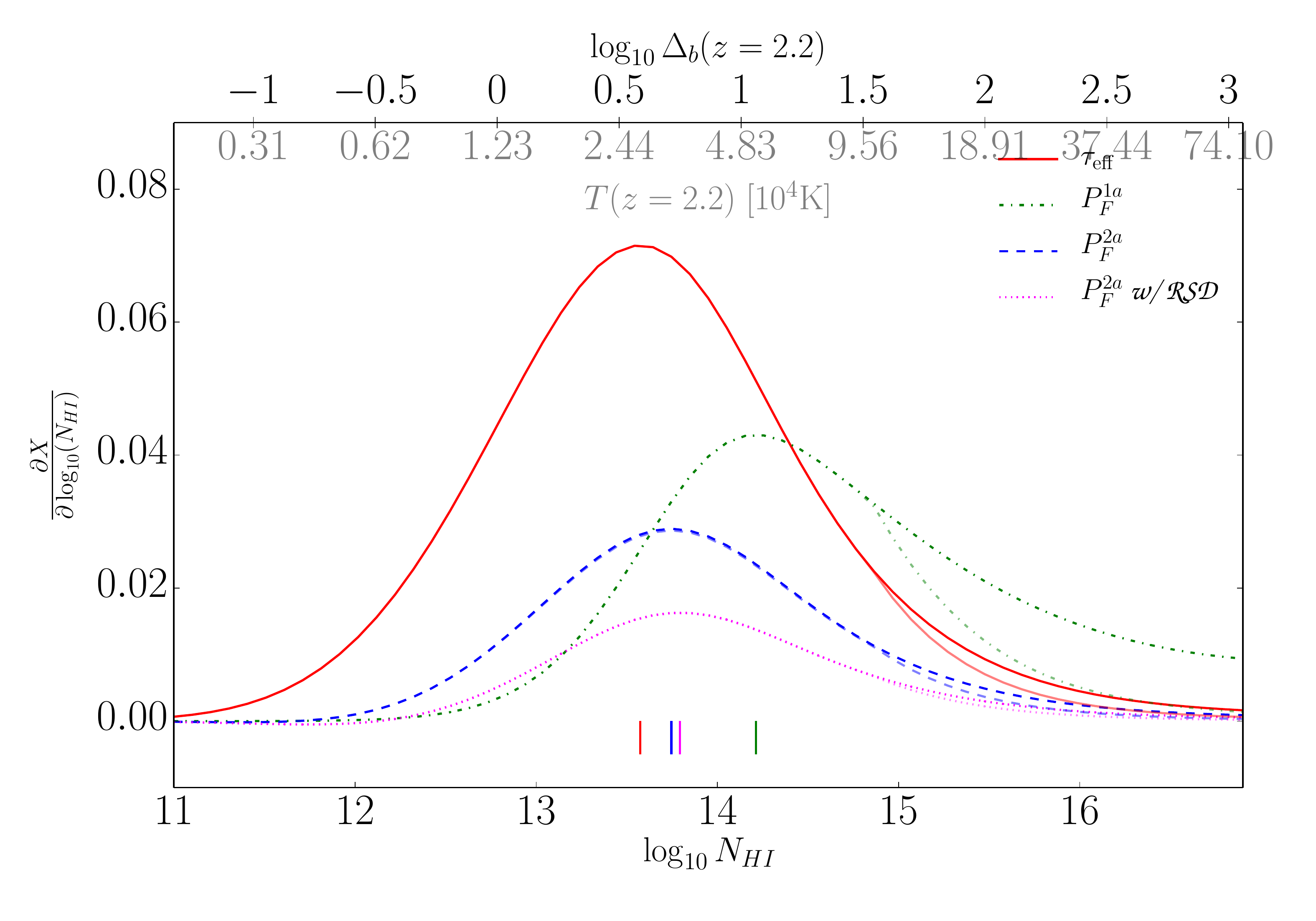}
  \caption{The differential of quantity $X$ with respect to column density,
    $\partial X/\partial \log_{10}(N_{HI})$, for $X = \tau_{\rm eff}$ (red solid
    curve), $X=P_F^{1a}$ (green dot-dashed curve), $X=P_F^{2a}$ (blue dashed
    curve), and $X=P_F^{2a}$ without redshift-space distortions (dotted
    magenta).  The vertical markers (with the same colour scheme) at the bottom
    of the plot indicate the maximum of the different distributions. These
    contributions to the power spectrum were computed at the pivot scale of
    $\sim 0.1\hMpc$.  The top axii show the gas density
    contrast $\Delta_b$ (above) and gas temperature $T_g$ (below) for our
    $z=2.2$ model. 
    The corresponding translucent curves are if we instead take an upper limit on the
    temperature of $10^5\;\mathrm{K}$, which mimics a more realistic thermal
    model.}
\label{fig:pdf_NHI}
\end{figure}

All Absorber Model quantities can be expressed as
\be
X = \int_{-\infty}^\infty \frac{\partial X}{\partial \log_{10}(N_{\rm HI})}
d\log_{10}(N_{\rm HI}),
\label{eq:pdf_X}
\ee 
where $X = \{ \tau_{\rm eff}, P_F^{1a}, P_F^{2a}, \dots \}$.  Thus, $\partial
X/\partial \log_{10}(N_{\rm HI})$ gives the contribution to $X$ per
$\log_{10}(N_{\rm HI})$. Fig.~\ref{fig:pdf_NHI} shows $\partial X/\partial
\log_{10}(N_{\rm HI})$ over the column density range probed by the
simulations. The flux power spectrum components depend also on the
wavenumber. Our calculations take $k = 10^{-3}\skm$, or $\approx 0.1\hMpc$,
at $z=2.2$, but the $X(N_{\rm HI})$ do not change significantly if $k$ is increased/decreased by a
factor of ten. Additionally, the top axis in Fig.~\ref{fig:pdf_NHI} relates the
column to the gas density in units of the mean, using the relation
from \cite{schaye01} that our model adopts.

The effective optical depth distribution, $\partial \tau_{\rm eff}/\partial
\log_{10}(N_{\rm HI})$, and the 2-absorber term, $\partial P_F^{2a}/\partial
\log_{10}(N_{\rm HI})$, peak at slightly lower columns of
$10^{13.6}\;\mathrm{cm^{-2}}$ than the 1-absorber term, $\partial
P_F^{1a}/\partial \log_{10}(N_{\rm HI})$, which peaks at
$10^{14.2}\;\mathrm{cm^{-2}}$. These trends are consistent with the previously
noted trends in the 1D power spectrum between different column density cuts
(Fig.~\ref{fig:pk_Ncuts}).  (We also find that redshift-space distortions act to
shift the peak of $\partial P_F^{2a}/\partial \log_{10}(N_{\rm HI})$ towards
lower columns.)  In terms of densities, at $z=2.2$ $\tau_{\rm eff}$ comes from a
the wide range of densities, $\Delta_b = 0.3 - 30$, with a peak at $\Delta_b
\approx 4$.  The range of densities affecting the 2-absorber signal is shifted
to somewhat higher densities, with a peak at $\Delta_b \approx 5$, and the
distribution is somewhat less broad ($\Delta_b = 0.7 - 10$). Finally, the
1-absorber term peaks at $\Delta_b \approx 10$, and most of its contribution
comes from between overdensities of several and halo-like overdensities.

The contribution for the 1-absorber term is skewed in the direction of higher
column densities. However, some of this skewness owes to unrealistically high
gas temperatures in the model from extrapolating a temperature density relation
to densities beyond where it applies (see top temperature axis in
Fig.~\ref{fig:pdf_NHI}), leading to unrealistically large linewidths.  The
translucent lines in Fig.~\ref{fig:pdf_NHI} show how the tail of the 1-absorber
contribution is diminished when the gas temperature is limited to be below
$10^5\;\mathrm{K}$, a limit that physically comes about from collisional cooling
\citep{2014MNRAS.444..503N, 2016MNRAS.456...47M}.\footnote{We have investigated
  more sophisticated models for temperature that follow the equilibrium
  temperature at high densities in the manner described in
  \cite{2014MNRAS.444..503N}, but find nearly identical results to this simple
  temperature-ceiling model.}  While other terms change negligibly, the Poisson
term now falls much faster towards zero contribution at higher
columns. Nevertheless, the 1-absorber term is sensitive to temperatures of the
gas at column densities of around $10^{14} - 10^{15}\;\mathrm{cm^{-2}}$, even if
we cap the temperature to mimic more realistic thermal physics. Moreover, with
the inclusion of column densities that are in the self-shielded regime (LLSs and
DLAs), the 1-absorber term's distribution likely develops a second peak at high
columns (when $f(N_{\rm HI})$ flattens out, and ${\mathrm EW}(N_{\rm HI})$
starts increasing with $\sim \sqrt{N_{\rm HI}}$ owing to damping wing
absorption).

For $z=3.0$, the same column density corresponds to a lower overdensity than at
$z=2.2$. Interestingly, we find that $\partial \tau_{\rm eff}/\partial
\log_{10}(N_{\rm HI})$ and $\partial P_F^{2a}/\partial \log_{10}(N_{\rm HI})$
peak at similar {\it gas overdensities} at $z=2.2$ and $z=3.0$, meaning that
these are shaped by similar gas densities but by higher column densities with
increasing redshift. Whereas, we find that $\partial P_F^{1a}/\partial
\log_{10}(N_{\rm HI})$ peaks at similar {\it column densities} at both
redshifts, meaning that this term is shaped by lower gas overdensities at higher
redshifts.

\section{Conclusion}
\label{sec:conclusion}
This paper presented a semi-analytic `Absorber Model' for the Ly$\alpha$ forest
that is inspired by the Halo Model.  This model is built on the absorption line
decomposition of the Ly$\alpha$ optical depth field, a decomposition that has
been used since the dawn of high-resolution Ly$\alpha$ forest observations.
This decomposition allows one to break up correlations into those within each
absorption system (`the 1-absorber term') and those between systems (`the
2-absorber term'), treating these systems as biased tracers of the underlying
matter fluctuations.  While the nonlinear exponential mapping between optical
depth and flux requires an infinite series of moments to calculate any
statistic, we show that this series can be re-summed (capturing the Poissonian
1-absorber term at all orders and truncating the 2-absorber at the desired order
in $\delta_L$).

We have focused on using this model to understand Ly$\alpha$ forest two-point
correlations, where we found:

\begin{itemize}
\item The modeled line-of-sight flux power spectrum (also known as the 1D power
  spectrum) agrees well with that in simulations.  For this comparison, we
  matched the column density distribution and linewidths in the model and the
  simulation mocks.  The primary additional ingredient that the model requires
  is the linear bias as a function of $N_{\rm HI}$, and we showed that a
  minimalist model predicts biases consistent with the simulations.  In
  addition, we showed that a simple linewidth model based on the model in
  \cite{garzilli15b} reproduced many aspects of the 1D power spectrum.  Absorber
  exclusion on the scale of the IGM Jeans length is needed to suppress the
  model's power by 30\% at intermediate wavenumbers to match the simulations.
\item The 1-absorber contribution to the 1D power spectrum is a factor of few
  times larger than the 2-absorber contribution (at least for the redshift range
  considered here of $z=2-3$).  The largeness of the 1-absorber term indicates
  that the 1D power at all wavenumbers is primarily shaped by the number of
  absorbers as a function of column.  To the extent that the sensitivity to
  cosmological parameters enters via the 2-absorber term, the 1-absorber term --
  which is a single number on scales much larger than the linewidths -- acts as
  a source of effective noise.  
   The dominance of Poisson fluctuations also
  helps explain why \citep{lai06,mcquinn11} found that large-scale temperature and intensity
  fluctuations have little effect on the 1D power spectrum.
\item We derived intuitive formulae for the effective optical depth and the
  $k\rightarrow 0$ limits of the 1- and 2-absorber terms, showing that they can
  be expressed simply as different equivalent-width weightings of the column
  density distribution (with an additional bias factor in the 2-absorber term's
  weighting).  At $z=2.2$, we found that the 2-absorber term is dominated by
  systems with $N_{\rm HI} \sim 10^{13}-10^{15}\;\mathrm{cm^{-2}}$, while the
  1-absorber term derives from those with $N_{\rm HI} \sim
  10^{14}-10^{15}\;\mathrm{cm^{-2}}$. Moreover, irrespective of redshift, the
  1-absorber contribution traces these same {\it column densities}, whereas
  the 2-absorber contribution traces the same {\it gas overdensities}, peaking
  at overdensities of $\sim 5$ and spanning $\sim 0.7-10$.
\item This is the first model that successfully predicts the Ly$\alpha$ forest
  linear biases in both density and velocity.  An objection to this claim might
  be that we have just redefined the uncertainty into $b(N_{\rm HI})$, but a
  counterargument is that $N_{\rm HI}$ tightly correlates with density
  \citep{schaye01, mcquinn11, altay11} and so a simple
  biasing model is possible.  Our simplest model for the linear velocity gradient bias,
  $b_{F, \eta}$, finds that it should equal $\tau_{\rm eff}$.  Our full model finds that peculiar velocity
  broadening contributes an additional 10-20\% correction that brings $b_{F, \eta}$
  into agreement with simulations. 
\end{itemize}

The Absorber Model has several deficiencies.  A major one is the unnatural
splitting of correlations into those within systems and between systems, a
problem that any halo-like model encounters.  Absorber exclusion is one
manifestation of this unnaturalness, which acts to suppress the 1-absorber term
on large-scales, but really appears in our 2-absorber term.  Another example is
the pressure smoothing of the gas, which both changes the profile of absorbers
(affecting mainly the 1-absorber term) and also likely alters the absorber
spacing (affecting the 2-absorber).  Our calculations mostly punted on modeling
these `filtering' effects.  Perhaps the most significant deficiency is that, in
contrast to the Halo Model, which is built upon the solid foundation of
excursion set theory \citep{1991ApJ...379..440B} and the attractor NFW halo
profile \citep{1996ApJ...462..563N}, the Absorber Model requires the column
density distribution (we took fits to it from simulations) as well as models for
the linewidth distribution and the line density bias.  The column density
distribution itself is ambiguously defined, as at some level the forest cannot
be decomposed into discrete systems.  This drawback should be especially
prescient at $z>4$, where much of the forest is saturated, and we expect the
Absorber Model will not be as successful.  On the positive side, the densities and
linewidths as a function of column have been studied and there is some analytic
understanding of both.  Our line model was able to build off this understanding to develop simple models for, e.g., the bias of lines.\\

There are many possible extensions to this model, including   
\begin{itemize}
\item Computing the model predictions at higher order in $\delta_L$ than second
  order, which in detail would require pursing a bias expansion for the
  absorbers \citep[e.g.][]{2017arXiv171106745D} or using some other model for
  the absorbers' nonlinear clustering.  (When the 2-absorber term is
  exponentiated, our formulae for moments of the flux field contain terms at
  every order in $\delta_L$, but they are only complete at second order.)  At
  second order in $\delta_L$, our model does not appear to capture the nonlinear
  scaling of the 3D flux power spectrum found in \cite{mcdonald03,arinyo15}.
  The 1-absorber term will likely be subdominant in the 3D power spectrum as it
  scales as $k^3$, and so the nonlinear scaling found in
  \cite{mcdonald03,arinyo15} likely owes to aspects of the nonlinear absorber
  clustering that are currently not included in our model.

\item Understanding how the shapes of correlations in the Ly$\alpha$ forest
  depend on the underlying cosmology. To the extent that the 2-absorber term
  dominates cosmological parameter constraints, the calculations presented here
  provide some understanding.  However, the small-scale width of absorption
  features should bear some cosmological dependence (that may be distinct from
  thermal effects), and the column density distribution also should depend on
  the cosmology.  These dependencies were not captured in our simplified
  calculations.

\item Including the damping wings of high-column density systems (Lyman-limit
  and damped Ly$\alpha$ systems).  This addition is a matter of simply including
  natural broadening in our linewidth model.  Several studies have shown that
  the effect of the wings on the power spectrum can be significant
  \cite{viel04DLA,font12b, mcdonald05b, mcquinnwhite, 2017arXiv170608532R}.
  This model could potentially provide a physically-motivated parametrization
  for their effect.

\item Including metal absorption that contaminates the Ly$\alpha$ forest.
  This could again be incorporated by altering the linewidth model to include
  absorption by various metal ions given some mapping between $N_{\rm HI}$ and
  the ions' optical depth profile.  A sophisticated model could assume some
  metallicity distribution as a function of column (as this has been constrained
  observationally \cite{2003ApJ...596..768S, 2004ApJ...602...38A,
    2008ApJ...689..851A}) and use photoionization modeling to include all
  potentially contaminating absorption lines.  Such a model for metal contamination could aid in understanding the imprints of metal contamination on
  the forest and how this contamination could bias parameter inferences.
  Similar calculations could be used to study the cross-correlation between the
  \lya\ forests and the absorption in other lines (such as the Ly$\beta$ or C IV
  forests).

\end{itemize}

State-of-the-art \lya\ forest studies rely heavily on cosmological hydrodynamic
simulations.  The Absorber Model is a complementary semi-analytic approach that
makes some analytic Ly$\alpha$ forest investigations possible.  Additionally, a
Monte-Carlo realization of this model could provide a more physical mock
Ly$\alpha$ absorption catalogue than the popular method that uses lognormal
transformations.

\section*{Acknowledgement}
 We thank Jamie Bolton for sharing a simulation from the Sherwood simulation
 suite.  This work is supported by US NSF grant AST-1514734, by the Institute
 for Advanced Study visiting faculty program, and by the Alfred P. Sloan
 foundation. The Sherwood simulations were performed on the Curie supercomputer,
 based in France at the Tres Grand Centre de Calcul (TGCC), using time awarded
 by the Partnership for Advanced Computing in Europe (PRACE) 8th call.  This
 Sherwood simulations also made use of the DiRAC High Performance Computing
 System (HPCS) and the COSMOS shared memory service at the University of
 Cambridge. These are operated on behalf of the Science and Technology
 Facilities Council (STFC) DiRAC HPC facility. This equipment is funded by BIS
 National E-infrastructure capital grant ST/J005673/1 and STFC grants
 ST/H008586/1, ST/K00333X/1.


\appendix
\section{Absorber model through Poisson distribution}
\label{app:poisson}
\label{sec:bias_beta}

Before we chose our discretization such that $p_i$ is either zero or one in a
cell.  Of course, to the extent that absorbers are uncorrelated (which is what
the Absorber Model assumes on small scales), $p_i$ is Poissonian, as the
absorbers are discrete.  Using that they are Poissonian allows us to re-sum all
Poissonian terms in the $\tau$ expansion expansion, as opposed to our approach
in the main text of keeping track of terms to third order in $\tau$ and then
motivating the resummation.  Keeping the notation adopted in
Section~\ref{sec:absorber_model}, we can rewrite the optical depth from
Eq.~\ref{eq:taux} such that each $\tau_i = \tau_a(x-x_i|N_i)$ in the sum over a
grid of $\Delta N$ and $\Delta x$ is multiplied by the occupation number of
elements ($n_i$), corresponding to $(N_i,x_i)$ within that volume
\be
\tau(x) = \sum_i n_i \tau_i.
\label{eq:taux_poisson}
\ee
The random variable $n_i$ is Poisson distributed with average
\be
\langle n_i \rangle \equiv {\bar n}(x_i|N_i) = f(N_i) \Big(1 + \delta_a(x_i|N_i) \Big) \Delta N
\Delta x,
\label{eq:poisson_mean}
\ee
where $\delta_a$ is the overdensity of absorbers.  [For linear configuration-space biasing,
$\delta_a(x_i|N_i) = b(N_i) \delta_L(x_i)$.]  Because $ \delta_a$ is not
Poissonian, this means that even though we have assumed $n_i$ is Poissonian in
each cell, the spatial distribution of absorbers is not owing to $\delta_a$.  We note that we are making the
same assumptions as in the main text, which follows the more standard approach.
With this, we can write the flux of one absorber at position $x$ as
\be
F(x) = e^{-\tau(x)} = e^{-\sum_i n_i \tau_i(x)} = \prod_i e^{-n_i \tau_i} \equiv
\prod_i F_i.
\ee

To
compute the mean of the flux requires averaging over the number the Poisson
distribution of the number densities as well as over $\delta_a$:
\be
\langle F \rangle = \left\langle \left\langle F(x) \right\rangle_n \right\rangle_\delta
 = \left\langle \left\langle \prod_i F_i \right\rangle_n \right\rangle_\delta =
 \left\langle \prod_i \left\langle F_i \right\rangle_n \right\rangle_\delta.
\label{eq:app:split}
\ee
The average over the Poisson distribution can be done analytically and yields
\begin{align}
\langle F \rangle = &\left\langle \prod_i \sum_{n_i=0}^{\infty} P(n_i|{\bar n}) e^{-
  n_i \tau_i} \right \rangle_\delta, \notag \\
=& \left\langle \prod_i e^{{\bar n}(x_i|N_i)
  \left(e^{-\tau_i(x)} - 1\right)}\right \rangle_\delta, \notag \\
=& \left \langle e^{\sum_i {\bar n}(x_i|N_i)
  \left(e^{-\tau_i(x)} - 1\right)}\right \rangle_\delta.
\end{align}

Using the expression for the average number density (Eq.~\ref{eq:poisson_mean}),
and again changing the sum $\sum_i \rightarrow \int \frac{dN_i dx_i}{\Delta N \Delta x}$ yields
\begin{align}
\langle F \rangle =& \left \langle e^{\int dN_1 \int dx_1 f_1 \left(1 + \delta_a(x_1) \right)
  \left(e^{-\tau_1(x)} - 1\right)} \right \rangle_\delta,
\label{eq:fake_F}
\end{align}
where we have replaced index $i$ with $1$ to match the notation in the main text (e.g. Eq.~\ref{eq:mean_F}). Just as in Sec.~\ref{sec:absorber_model}, using the cumulant theorem we compute the
$\delta$--average, which gives to quadratic order in $\delta_L$:
\begin{align}
\langle F \rangle =& \exp\left[\int_N d\vN \int_x d\vx f_1
  \left(e^{-\tau_1(x)} - 1\right) + \right. \notag \\
+&\left.\phantom{\int} \frac{1}{2}\int_N d^2\vN \int_x d^2\vx
  f_1 f_2 \xi^L_{12}(x_2 - x_1)\left(e^{-\tau_1(x)} - 1\right)\left(e^{-\tau_2(x)} - 1\right) \right],
\end{align}
where as in the main text $\langle \delta_a(x_1) \delta_a(x_2) \rangle = \xi_{12}^L(x_2 -
x_1)$. With the replacement $\tau_1(x) \rightarrow \tau_1(y)+\tau_1(z)$, the above expression then becomes
$\langle F(y) F(z) \rangle$, just as in Sec.~\ref{sec:absorber_model}.

Because this approach yields an expression for the flux given the large-scale
density, $\delta_a$ (the term in the expectation value of Eq.~\ref{eq:fake_F}), it
nicely allows one to compute the flux biases of the density ($b_{F,\delta}$) and
of the velocity gradient ($b_{F,\eta}$). The density bias of the flux is defined
as
\be
b_{F,\delta}(k) \equiv \left\langle \frac{\partial \delta_F(k)}{\partial
  \delta_L(k)} \right\rangle = 
\frac{1}{\langle F \rangle} \left\langle \frac{\partial F(k)}{\partial \delta_L(k)} \right\rangle,
\ee
where the (constant) linear bias, which we denote as $b_{F,\delta}$ in the main
text, is obtained by taking the ${k \rightarrow 0}$ limit. Using the above
notation, $b_{F,\delta}(k) $ can be written as
\begin{align}
b_{F,\delta}(k) &= \frac{1}{\langle F \rangle} \left\langle\left\langle \frac{\partial F(k)}{\partial
  \delta_L(k)} \right\rangle_n \right\rangle_\delta, \notag \\
&= \frac{1}{\langle F \rangle} \left\langle \frac{\partial}{\partial
  \delta_L(k)} \left\langle F(k)\right\rangle_n \right\rangle_\delta,
\notag \\
&= \frac{1}{\langle F \rangle} \left\langle \frac{\partial}{\partial
  \delta_L(k)} \int dx e^{-ikx} e^{\int dN_1 f_1 \int dx_1 \left(1+b_1
  \delta_L(x_1) + {\cal O}(\delta_L^2)\right)\left( e^{-\tau_1(x)} -1 \right)}
\right\rangle_\delta,
\end{align}
where for simplicity we have ignored redshift-space distortions in our
expression as they do not alter $b_{F,\delta}(k)$.

To simplify further, we drop terms of ${\cal O}(\delta_L^2)$ as these will yield
zero in the desired $k\rightarrow0$ limit (with a caveat discussed below) and consider a finite interval $L$ so
we can write $\delta_L(x) = \sum_{n=-\infty}^\infty \delta_L(k) e^{i k x} \Delta
k/2\pi$ where $\Delta k = 2\pi/L$ and $k = n \Delta k$ for integer $n$.  Then,
rewriting the previous equation yields
\begin{align}
b_{F,\delta}(k) &= \frac{1}{\langle F
  \rangle} \left\langle \frac{\partial}{\partial
  \delta_L(k)} \int dx e^{-ikx} e^{-\int dN_1 \int dx_1 f_1 \left(1 + b_1
   L^{-1} \sum \delta_L(k') e^{i k' x_1} \right)
  \left(1 - e^{-\tau_1(x)}\right)} \right\rangle_\delta, \notag \\
&= \frac{-1}{\langle F
  \rangle} \left\langle   \int dx e^{-ikx} e^{\int dN_1 f_1 \int dx_1 \left(1+b_1
  \delta_L(x_1)\right)\left( e^{-\tau_1(x)} -1 \right)}  L^{-1} \int dN_1 \int dx_1 f_1 b_1
   {K}_1(x) e^{i k x_1} \right\rangle_\delta,
  \notag \\
&= -  L^{-1} \int dx \int dN_1 \int dx_1 f_1 b_1
   {K}(x -x_1) e^{-i k (x- x_1)},
  \notag \\
&= -\int dN_1 f_1 b_1 {\widetilde K}_1(k),\label{eq:bfd2}
\end{align}
where ${\widetilde K}_1(k)$ is the Fourier transform of the kernel $K_1(x)$.
Note that all spatial integrals are over range $-L/2$ to $L/2$.  Going from the
first line to the second has the derivative hit the mode with $k' = k$.
(Additionally, we note that our previous dropping of the ${\cal O}(\delta_L^2)$
terms, results in $\delta_L$ only being in the exponent after the derivative.
This simplification is not necessarily justified, but in our case is because we
find the clustering component to $\tau_{\rm eff}$ [the term we are dropping] is
negligible.  See the discussion following Eq.~\ref{eq:mF-simple}.) Going from
the second to the third recognizes that the $\langle...\rangle_\delta$ can be
brought in to average over a quantity that is equal to $\langle F \rangle$.  The
third to final line notices that the $x_1$ integral over $K$ is a Fourier
transform and that the integrand does not depend on $x$.  Eq.~\ref{eq:bfd2} is
identical to Eq.~\ref{eqn:bFd}, where we had extracted $b_{F,\delta}$ from the
convolution terms in the flux correlation function.

Following nearly identical steps as above, and noting that on large-scales $\delta_a
=(1 + b_1 \delta_L) /(1 - \eta_L)$,\footnote{Just like for $\delta_L$, higher
  order in $\eta_L$ will not affect the $k\rightarrow 0$ results.} the velocity
gradient bias can be shown to be
\begin{align}
b_{F,\eta}(k) &\equiv \frac{1}{\langle F
  \rangle} \left\langle  \frac{\partial F(k)}{\partial
  \eta_L(k)} \right\rangle
= - \int dN_1 f_1 {\widetilde K}_1(k).
\label{eq:b_eta}
\end{align}
The flux bias of the velocity gradient is, to linear order, of the same absolute
value as the effective optical depth ($\tau_{\rm eff}$).

However, the above biases assumed that the line profiles, that enter in $K$, do
not depend on $\delta_L$ or $\eta_L$.  The linewidths likely depend on
$\eta_L$, which Sec.~\ref{sec:simple_bias} motivated is necessary to explain
$b_{F,\eta}$ in both the simulations and observations.  This extension yields
Eq.~\ref{eqn:bfeta} in the main text for $b_{F,\eta}$, the $k\rightarrow 0$
limit of $b_{F,\eta}(k)$, noting that ${\widetilde K}_1(k=0) = {\rm
  EW}$. Appendix~\ref{ss:vgblm} calculates this for a simple extension of our
line model (finding results that agree well with simulations and observations).

\section{Linewidth model}
\label{sec:thermal_model}

One of the inputs of the framework presented in Sec.~\ref{sec:absorber_model} is
the dependence of the absorption linewidths on the column density. This paper
adopts a simple model, based on \cite{Garzilli15}.  See \cite{Garzilli15} for plots demonstrating the success of this model. The model proposes that the
column density of a given absorption system ($N_{\rm HI}$) is proportional to
the neutral hydrogen number density at that location ($n_{\rm HI}$), times the
characteristic length scale, associated with the size of the absorbers. Taking
the length scale to be the Jeans scale ($\lambda_J$), which is supported by both
simple arguments \cite{schaye01} and complex cosmological simulations \cite{mcquinn11, altay11},
\be
N_{\rm HI} = f_N \lambda_J n_{\rm HI} \propto \Delta_b^{\frac{3}{2}} T^{-0.76},
\ee
where $f_N$ is an order unity fudge factor in the relation. In this paper, we
choose $f_N$ to be a constant with value of $0.3$ as motivated in
\cite{Garzilli15}.  The number density of neutral hydrogen ($n_{\rm HI}$) is
related to the number density of hydrogen nuclei ($n_{\rm H}$) through the usual
photoionization equilibrium relation
\be
\frac{n_{\rm HI}}{n_{\rm H}} = \alpha_{\rm HI,\gamma}(T) \Gamma_{\rm HI,\gamma}^{-1}
n_e,
\ee
where the photo-ionization rate ($\Gamma_{\rm HI,\gamma}$) is considered to be
spatially uniform and the recombination rate ($\alpha_{\rm HI,\gamma}$) is
approximated as $\alpha_{\rm HI,\gamma} = \alpha_0 T_{4}^{-0.76}$, with
$\alpha_0 = 4.3 \times 10^{-13} \;\mathrm{cm^3 s^{-1}}$ and $T_4 =
T/10^4\;\mathrm{K}$. The electron number density for optically thin gas can be
related to the hydrogen number density as $n_e = n_H (1 - Y/2)/(1-Y)$, where
$Y=0.24$ is the helium (He) mass fraction and we have assumed that the helium is
doubly ionized (as is applicable at $z\lesssim 3$). Furthermore, the hydrogen
number density is related to the total mass density as
\be
\rho_m = \frac{n_H m_{\rm H}}{f_g (1-Y)},
\ee
where $m_{\rm H}$ is the mass of the hydrogen atom, and $f_g =
\Omega_b/\Omega_m$ is the gas fraction. Finally, using the standard power-law
parameterization of temperature and density,
\be
T = T_0 \Delta_b^{\gamma-1},
\ee
a one-to-one correspondence between the column density and the (non-linear)
over-density can be obtained by combining the above equations.  We use this
relation in what follows to calculate $\lambda_J$ (as well as the bias of lines
in Appendix C).  

With these basics out of the way, we can now write expressions for the $\sigma$ of our model's Gaussian-in-optical-depth absorption lines:
\be
\tau(x|\sigma) = \sigma_0 N \frac{1}{\sigma \sqrt{\pi}} e^{-x^2/\sigma^2}.
\ee
 
 Some of the broadening owes to
from thermal Doppler broadening. In this case, the width of the lines can be
described as
\be
\sigma_D^2 = \frac{2 k_B T(N_{\rm HI})}{m_{\rm H}},
\ee
where $k_B$ is the Boltzmann constant and $\sigma_D$ has units of velocity.
Following \cite{Garzilli15}, in addition to Doppler broadening we also include
the physical velocity broadening, making the ansatz that this broadening is
dominated by the Hubble flow rather than the peculiar velocity (which is more
apt at low densities). Thus, we write
\be
\sigma_H = f_N \lambda_J H(z),
\label{eq:hubble_broadening}
\ee
where $H(z)$ is the Hubble function.\footnote{Unlike in \cite{Garzilli15}, we have here
included the same factor $f_N$ that should characterize the distribution nature
of the typical size of the absorbers around the Jeans length value, although
this change has a minimal impact on our results.} The relation for the Jeans length used in the paper is
\be
\lambda_J^2 = \frac{c_s^2 \pi}{G\rho_m} = \frac{5 k_B T}{3 \mu m_{\rm H}}
\frac{\pi}{G \rho_m},
\label{eq:lJ}
\ee
where $G$ is the gravitational constant, $c_s$ is the sound speed, and
$\mu=0.59$ is the mean molecular weight for ionized primordial gas. 

Combining the results we can write the total linewidth of the absorbers as
\be
\sigma^2 = \sigma_D^2 + \sigma_H^2 = \sigma_D^2 \left( 1 + \frac{5}{9\mu} \Omega_m(z)^{-1}
\frac{(2\pi f_N)^2}{\Delta_b} \right),
\ee
where $\Omega_m(z)$ is the coeval matter energy density.  

Even though the model is simplistic in that it does not capture the scatter
found in hydrodynamical simulations or the contribution to broadening from
peculiar velocities, the main text presents results in a way that tries to be as
model independent as possible (see Section \ref{sec:mocks}). 

\subsection{Velocity gradient bias in this linewidth model}
\label{ss:vgblm}

We would like to estimate the amount $b_{F,\eta}$ is affected by the broadening of lines in response to a large-scale velocity gradient.  Using the Hubble
broadening model just described, we can
add the inhomogeneity to Eq.~\ref{eq:hubble_broadening}, saying that
the Gaussian profiles have an additional physical velocity
broadening component such that\footnote{Eq.~\ref{eq:dv} uses that the velocity across an absorber is
  \be
    \Delta v \equiv v(x_2) - v(x_1) = a(z) H(z) (x_2 - x_1) + v_p(x_1)
    - v_p(x_2) \approx a(z) H(z) L \left(1 + \frac{1}{aH}\frac{\partial
      v_p}{\partial x}\right) = a(z) H(z) L (1- \eta),
      \label{eq:dv}
  \ee
  where the size of the absorber is $L = x_2 - x_1$.
} 
\be
\sigma_H = f_N \lambda_J H(z) (1 - \eta_L).
\ee
The full expression for the linewidth from velocity effects, $\sigma_H$, should have the nonlinear $\eta$ rather than its linear theory limit $\eta_L$, as we have written, where $\eta \equiv -\partial v_p/\partial x / (aH)$.  We make the ansatz here that we can replace $\eta$ with $\eta_L$.  We can now evaluate $d\log {\rm EW}/d\eta_L$ in our full expression for the velocity gradient bias (Eq.~\ref{eqn:bfeta}), finding

\be
b_{F,\eta} = - \int dN f(N) {\rm EW} \left(  1 - \alpha^{(\eta)} \frac{\partial
  {\ln{\rm EW}}}{\partial \ln{\sigma}}  \right),
\ee
where $\alpha^{(\eta)} = 1/[1 + 9\mu \Delta_b(N) \Omega_m(z)/(2\sqrt{5}\pi
f_N)^2]$ and $\Delta_b(N)$ is the nonlinear density contrast (its dependence on the column density is given by the model described
in Appendix~\ref{sec:thermal_model}). 

In this simple extension of the linewidth model, the additional ${\partial
  {\ln{\rm EW}}}/{\partial \ln{\sigma}}$ term leads to small deviations from unity of the
$b_{\tau,\eta} = -{b_{F,\eta}}/{\ln{\langle F \rangle}}$.  At $z=2.2$, the value of $b_{\tau,\eta}$ is $0.89$ in our model, and at $z=3.0$ this becomes $0.82$. Both of these points agree well
with the predictions of the simulations \citep{arinyo15}.

\section{Simple bias model}
\label{sec:bias}

Our simple bias model assumes that there is a one-to-one relation between
$N_{\rm HI}, \Delta_b$ and $\delta_L$ smoothed on the filtering scale where the
density variance is $\sigma_f^2$. We will also assume that the mass of the
systems is the Jeans mass for $\Delta_b$, which is supported by cosmological
simulations \citep{schaye01, mcquinn11, altay11}.

For this setup, we relate the number of systems that reside at a given column to the
number of systems that stem from a given linear density
\be
\frac{d^2{\cal N}}{dx dN_{\rm HI}} = \frac{d^2 {\cal N}}{dx d\delta_L}
\frac{d\delta_L}{d\Delta_{b}} \frac{d\Delta_{b}}{dN_{\rm HI}}.
\ee
We can also write an expression for the 3D density of absorbers that
start from linear density in the range $d\delta_L$ and that fragment into
objects of size of the Jeans mass, $M_J(\Delta_{b})$:
\be
\frac{d^4{\cal N}}{d^3{\vec x} d\delta_L} = \frac{\bar
  \rho}{M_J(\Delta_{b})} \frac{1}{\sqrt{2\pi \sigma_f^2}} \exp\left[-\frac{\delta_L^2}{2\sigma_f^2}\right],
\ee
where $\sigma_f$ is the characteristic variance in $\delta_L$ for a region of size that maps into an absorbers, taken to be the filtering length (proportional to the Jeans length at the mean density\cite{gnedin98}),\footnote{There is some inconsistency inherent in this model as we require a single scale to measure $\sigma_f$ but allow $M_J$ to be a function of density. This inconsistency is lessened by $\sigma_f$ being a weak function of scale.}
and finally
\be
\frac{d^2{\cal N}}{dx d\delta_L} = \lambda_J^2\frac{d^4{\cal N}}{d^3{\vec x} d\delta_L},
\ee
where we have taken the Jeans length squared, $\lambda_J^2$, to be the characteristic
cross section of the system, noting $M_J = {\bar \rho} (1+\Delta_{b})
\lambda_J^3$. Since this is a crude model, we drop the typical factors of $4\pi/3$ and $\pi$ in the
relations.  Given an empirical $f(N_{\rm HI})$, the previous allows us to relate
$N_{\rm HI}$ to $\delta_L$:
\be
\frac{1}{\cal M}\left[\int_{N_{\rm HI}}^{\infty} (1+\delta_{\rm NL}) \lambda_J(\delta_{\rm NL})
f(N_{\rm HI}) dN_{\rm HI} +\varepsilon_>\right]= \int_{\delta_L}^{\infty} \frac{1}{\sqrt{2\pi
    \sigma_f^2}} \exp\left[-\frac{\delta_L^2}{2\sigma_f^2}\right]
d\delta_L,
\label{eq:deltaL_relation}
\ee
where  
\be
{\cal M}(\varepsilon_<, \varepsilon_>) = \varepsilon_<  + \int_{0}^{\infty} (1+\delta_{\rm NL}) \lambda_J(\delta_{\rm NL})
f(N_{\rm HI}) dN_{\rm HI} + \varepsilon_>.
\label{eq:norm}
\ee
In the case of $\varepsilon_< = \varepsilon_> = 0$, Eq.~\ref{eq:deltaL_relation}
is just integrating $M_J \frac{d^2{\cal N}}{dx^3 d\delta_L} d\delta_L = M_J
\frac{d^2{\cal N}}{dx^3 dN_{\rm HI}} dN_{\rm HI}$.  Solving this case generates
a one-to-one and onto mapping between $N_{\rm HI}$ and $\delta_L$.

However, as discussed in the Section~\ref{ss:modelprops}, line fitting misses some of the
least- and most-dense gas.  Our model captures the missing diffuse missing component with
$\varepsilon_<$, as $\varepsilon_<$ identifies $N_{\rm HI} =0$ with the
$\delta_L$ that is larger than all but $\approx \varepsilon_</{\cal M}$ of the
possible $\delta_L$.  While we expect $\varepsilon_<\lesssim 0.1$, we show below that the precise value of $\varepsilon_<$ is
not crucial (as anything finite and small yields roughly the same result).  Even more important, not all the gas in
the Universe is visible in the forest.  Tens of percent are shock heated at the
redshifts of interest, primarily within galactic dark matter halos.
Eq.~\ref{eq:norm} identifies $N_{\rm HI} =\infty$ (in actuality, $N_{\rm HI}
=10^{17}$cm$^{-2}$ is the maximum in our calculations) with the $\delta_L$ that
is only smaller than $\varepsilon_>/{\cal M}$ of all other $\delta_L$.  Thus, if a
quarter of gas is shock heated and not visible at $z=2.2$ (as simulations find),
and this shock heated gas is identified with the largest $\delta_L$, we expect
$\varepsilon_> \sim 0.5$. This estimate uses that ${\cal M}(0,0) = 1.44$ at $
z=2.2$ (${\cal M}(0,0) = 1.36$ at $z=3.0$) in our calculations.  This estimate
for $\varepsilon_>$ is roughly consistent with what we find is needed to match
the large-scale bias of the {\it mocks}, especially when accounting for line
exclusion.  Including $(\varepsilon_<, \varepsilon_>)$ in this bias model is important; otherwise our lowest and highest columns have divergent negative and positive biases, respectively.

This mapping between $N_{\rm HI}$ and $\delta_L$ is the most important part of
the bias model as the Lagrangian bias follows:
\be
b_{N_{\rm HI}} = \lim_{\delta_{L,{\rm lw}}\rightarrow 0, \sigma_{\rm lw}^2\rightarrow 0}
\frac{d\log{n(N_{\rm HI}|\delta_{L,{\rm lw}})}}{d\delta_{L,{\rm lw}}} =
\frac{\delta_L(N_{\rm HI})}{\sigma_f^2},
\ee
where $\delta_{L,{\rm lw}}$ is the long wavelength fluctuation with variance
$\sigma_{\rm lw}^2$.  The final equality uses that
\be
n(N_{\rm HI}|\delta_{L,{\rm lw}}) \equiv \frac{d^4{\cal N}}{d^3{\vec x}
  d\delta_L} (\delta_L|\delta_{L,{\rm lw}}) = \frac{\bar
  \rho}{M_J(\delta_{\rm NL})} \frac{1}{\sqrt{2\pi (\sigma_f^2-\sigma_{\rm lw}^2)}} \exp\left[-\frac{(\delta_L-\delta_{L,{\rm lw}})^2}{2(\sigma_f^2-\sigma_{\rm lw}^2)}\right].
\ee
We can now express the bias in terms of $N_{\rm HI}$ using
Eq.~\ref{eq:deltaL_relation} for $\delta_L(N_{\rm HI})$.
We fix the value of the filtering scale to $k_F^{-1} = 41\;\mathrm{c kpc/h}$
(taking it to be $2.2$ times the Jeans wavenumber, c.f. Eq.~\ref{eq:lJ}, at
$\Delta_b=1$ and $z=2.2$ as motivated by \cite{gnedin98}) such that the standard
deviation in a region of size the filtering scale is $\sigma_f = 1.68$.

\begin{figure}[!h]
  \centering
  \includegraphics[width=1.0\linewidth]{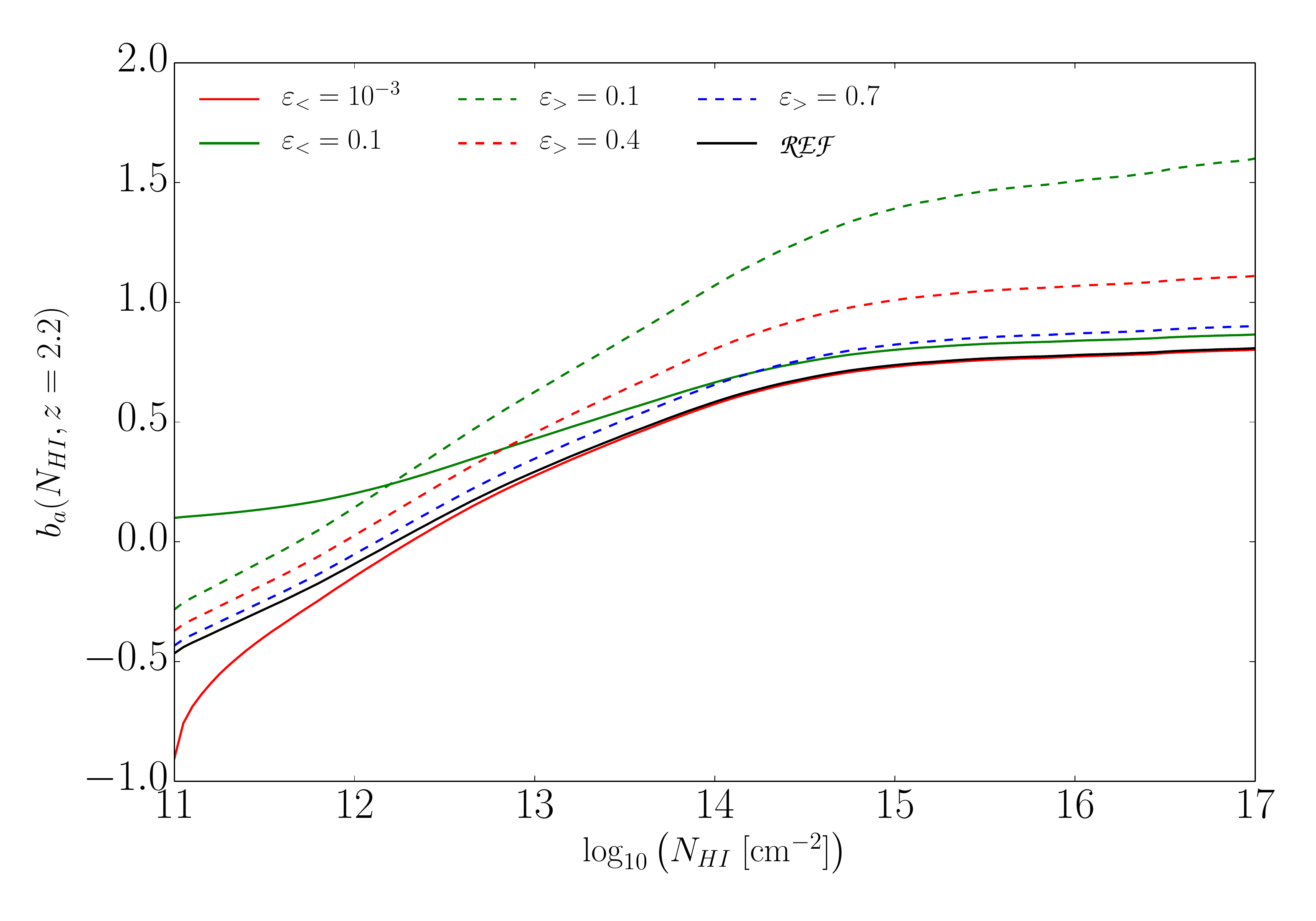}
  \caption{Column density dependence of our simple bias model at $z=2.2$ for
    different values of $(\varepsilon_<,\varepsilon_>)$. The black lines show
    the Reference Model of $(\varepsilon_<,\varepsilon_> ) = (0.01, 0.9)$ and
    $\sigma_f= 1.68$. The array of
    solid colored lines vary $\varepsilon_<$ from the refernece; the array of
    dashed colored lines vary $\varepsilon_>$.}
\label{fig:bias_epsilon}
\end{figure}

Figure ~\ref{fig:bias_epsilon} shows the effect of varying the parameters
$(\varepsilon_<,\varepsilon_>)$ on $b(N_{\rm HI})$. Observations suggest that
for DLAs, i.e. $N_{\rm HI} > 10^{20.3}$cm$^{-2}$, $b(N_{\rm HI}) \approx 2$
\citep{font12c,perez17}, so the bias values should be below this threshold at
the lower $N_{\rm HI}$ we consider.\footnote{For some of our models $b(N_{\rm
    HI}) \approx 1$ at the highest columns we consider.  We do not think such
  biases are inconsistent with the larger $b(N_{\rm HI}) \approx 2$ of DLAs as
  bias increases quickly for the rarest systems.} The exact values of the
$(\varepsilon_<,\varepsilon_>)$ parameters for the Reference Model were fixed
such that the low-$k$ power spectrum agrees with that of the {\it mocks}.  The
values required for $(\varepsilon_<,\varepsilon_>)$ in the Reference Model as
well as the other cases shown in Figure ~\ref{fig:bias_epsilon} are consistent
with the range we physically motivate, except $\varepsilon_> = 0.1$ which is
outside this range.  Especially if we exclude this $\varepsilon_> = 0.1$ curve,
the range of bias values of the different models is relatively small.  It would
be interesting to compare the bias values in this model to full numerical
simulations

\bibliographystyle{JHEP}
\bibliography{Bibliofile}

\providecommand{\href}[2]{#2}\begingroup\raggedright\begin{thebibliography}{10}

\bibitem{mcdonald00}
P.~{McDonald}, J.~{Miralda-Escud{\'e}}, M.~{Rauch}, W.~L.~W. {Sargent}, T.~A.
  {Barlow}, R.~{Cen}, and J.~P. {Ostriker}, {\it {The Observed Probability
  Distribution Function, Power Spectrum, and Correlation Function of the
  Transmitted Flux in the Ly{$\alpha$} Forest}},  {\em \apj} {\bf 543} (Nov.,
  2000) 1--23, [\href{http://arxiv.org/abs/astro-ph/9911196}{{\tt
  astro-ph/9911196}}].

\bibitem{zaldarriaga01}
M.~{Zaldarriaga}, U.~{Seljak}, and L.~{Hui}, {\it {Correlations in the
  Ly{$\alpha$} Forest: Testing the Gravitational Instability Paradigm}},  {\em
  \apj} {\bf 551} (Apr., 2001) 48--56,
  [\href{http://arxiv.org/abs/astro-ph/0}{{\tt astro-ph/0}}].

\bibitem{croft02}
R.~A.~C. {Croft}, D.~H. {Weinberg}, M.~{Bolte}, S.~{Burles}, L.~{Hernquist},
  N.~{Katz}, D.~{Kirkman}, and D.~{Tytler}, {\it {Toward a Precise Measurement
  of Matter Clustering: Ly{$\alpha$} Forest Data at Redshifts 2-4}},  {\em
  \apj} {\bf 581} (Dec., 2002) 20--52,
  [\href{http://arxiv.org/abs/astro-ph/0012324}{{\tt astro-ph/0012324}}].

\bibitem{zaldarriaga03}
M.~{Zaldarriaga}, R.~{Scoccimarro}, and L.~{Hui}, {\it {Inferring the Linear
  Power Spectrum from the Ly{$\alpha$} Forest}},  {\em \apj} {\bf 590} (June,
  2003) 1--7, [\href{http://arxiv.org/abs/astro-ph/0}{{\tt astro-ph/0}}].

\bibitem{seljak03}
U.~{Seljak}, P.~{McDonald}, and A.~{Makarov}, {\it {Cosmological constraints
  from the cosmic microwave background and Lyman {$\alpha$} forest revisited}},
   {\em \mnras} {\bf 342} (July, 2003) L79--L84,
  [\href{http://arxiv.org/abs/astro-ph/0}{{\tt astro-ph/0}}].

\bibitem{mcdonald03}
P.~{McDonald}, {\it {Toward a Measurement of the Cosmological Geometry at z
  \~{} 2: Predicting Ly{$\alpha$} Forest Correlation in Three Dimensions and
  the Potential of Future Data Sets}},  {\em \apj} {\bf 585} (Mar., 2003)
  34--51, [\href{http://arxiv.org/abs/astro-ph/0108064}{{\tt
  astro-ph/0108064}}].

\bibitem{viel04}
M.~{Viel}, M.~G. {Haehnelt}, and V.~{Springel}, {\it {Inferring the dark matter
  power spectrum from the Lyman {$\alpha$} forest in high-resolution QSO
  absorption spectra}},  {\em \mnras} {\bf 354} (Nov., 2004) 684--694,
  [\href{http://arxiv.org/abs/astro-ph/0404600}{{\tt astro-ph/0404600}}].

\bibitem{viel04bis}
M.~{Viel}, S.~{Matarrese}, A.~{Heavens}, M.~G. {Haehnelt}, T.-S. {Kim},
  V.~{Springel}, and L.~{Hernquist}, {\it {The bispectrum of the Lyman
  {$\alpha$} forest at z\~{} 2-2.4 from a large sample of UVES QSO absorption
  spectra (LUQAS)}},  {\em \mnras} {\bf 347} (Jan., 2004) L26--L30,
  [\href{http://arxiv.org/abs/astro-ph/0308151}{{\tt astro-ph/0308151}}].

\bibitem{viel04hrwmap}
M.~{Viel}, J.~{Weller}, and M.~G. {Haehnelt}, {\it {Constraints on the
  primordial power spectrum from high-resolution Lyman {$\alpha$} forest
  spectra and WMAP}},  {\em \mnras} {\bf 355} (Dec., 2004) L23--L28,
  [\href{http://arxiv.org/abs/astro-ph/0}{{\tt astro-ph/0}}].

\bibitem{mcdonald05}
P.~{McDonald}, U.~{Seljak}, R.~{Cen}, D.~{Shih}, D.~H. {Weinberg}, S.~{Burles},
  D.~P. {Schneider}, D.~J. {Schlegel}, N.~A. {Bahcall}, J.~W. {Briggs},
  J.~{Brinkmann}, M.~{Fukugita}, {\v Z}.~{Ivezi{\'c}}, S.~{Kent}, and D.~E.
  {Vanden Berk}, {\it {The Linear Theory Power Spectrum from the Ly{$\alpha$}
  Forest in the Sloan Digital Sky Survey}},  {\em \apj} {\bf 635} (Dec., 2005)
  761--783, [\href{http://arxiv.org/abs/astro-ph/0407377}{{\tt
  astro-ph/0407377}}].

\bibitem{mcdonald06}
P.~{McDonald}, U.~{Seljak}, S.~{Burles}, D.~J. {Schlegel}, D.~H. {Weinberg},
  R.~{Cen}, D.~{Shih}, J.~{Schaye}, D.~P. {Schneider}, N.~A. {Bahcall}, J.~W.
  {Briggs}, J.~{Brinkmann}, R.~J. {Brunner}, M.~{Fukugita}, J.~E. {Gunn}, {\v
  Z}.~{Ivezi{\'c}}, S.~{Kent}, R.~H. {Lupton}, and D.~E. {Vanden Berk}, {\it
  {The Ly{$\alpha$} Forest Power Spectrum from the Sloan Digital Sky Survey}},
  {\em \apjs} {\bf 163} (Mar., 2006) 80--109,
  [\href{http://arxiv.org/abs/astro-ph/0405013}{{\tt astro-ph/0405013}}].

\bibitem{seljak06}
U.~{Seljak}, A.~{Slosar}, and P.~{McDonald}, {\it {Cosmological parameters from
  combining the Lyman-{$\alpha$} forest with CMB, galaxy clustering and SN
  constraints}},  {\em \jcap} {\bf 10} (Oct., 2006) 14,
  [\href{http://arxiv.org/abs/astro-ph/0604335}{{\tt astro-ph/0604335}}].

\bibitem{slosar11}
A.~{Slosar}, A.~{Font-Ribera}, M.~M. {Pieri}, J.~{Rich}, J.-M. {Le Goff},
  {\'E}.~{Aubourg}, J.~{Brinkmann}, N.~{Busca}, B.~{Carithers},
  R.~{Charlassier}, M.~{Cort{\^e}s}, R.~{Croft}, K.~S. {Dawson},
  D.~{Eisenstein}, J.-C. {Hamilton}, S.~{Ho}, K.-G. {Lee}, R.~{Lupton},
  P.~{McDonald}, B.~{Medolin}, D.~{Muna}, J.~{Miralda-Escud{\'e}}, A.~D.
  {Myers}, R.~C. {Nichol}, N.~{Palanque-Delabrouille}, I.~{P{\^a}ris},
  P.~{Petitjean}, Y.~{Pi{\v s}kur}, E.~{Rollinde}, N.~P. {Ross}, D.~J.
  {Schlegel}, D.~P. {Schneider}, E.~{Sheldon}, B.~A. {Weaver}, D.~H.
  {Weinberg}, C.~{Yeche}, and D.~G. {York}, {\it {The Lyman-{$\alpha$} forest
  in three dimensions: measurements of large scale flux correlations from BOSS
  1st-year data}},  {\em \jcap} {\bf 9} (Sept., 2011) 1,
  [\href{http://arxiv.org/abs/1104.5244}{{\tt arXiv:1104.5244}}].

\bibitem{busca13}
N.~G. {Busca}, T.~{Delubac}, J.~{Rich}, S.~{Bailey}, A.~{Font-Ribera},
  D.~{Kirkby}, J.-M. {Le Goff}, M.~M. {Pieri}, A.~{Slosar}, {\'E}.~{Aubourg},
  J.~E. {Bautista}, D.~{Bizyaev}, M.~{Blomqvist}, A.~S. {Bolton}, J.~{Bovy},
  H.~{Brewington}, A.~{Borde}, J.~{Brinkmann}, B.~{Carithers}, R.~A.~C.
  {Croft}, K.~S. {Dawson}, G.~{Ebelke}, D.~J. {Eisenstein}, J.-C. {Hamilton},
  S.~{Ho}, D.~W. {Hogg}, K.~{Honscheid}, K.-G. {Lee}, B.~{Lundgren},
  E.~{Malanushenko}, V.~{Malanushenko}, D.~{Margala}, C.~{Maraston},
  K.~{Mehta}, J.~{Miralda-Escud{\'e}}, A.~D. {Myers}, R.~C. {Nichol},
  P.~{Noterdaeme}, M.~D. {Olmstead}, D.~{Oravetz}, N.~{Palanque-Delabrouille},
  K.~{Pan}, I.~{P{\^a}ris}, W.~J. {Percival}, P.~{Petitjean}, N.~A. {Roe},
  E.~{Rollinde}, N.~P. {Ross}, G.~{Rossi}, D.~J. {Schlegel}, D.~P. {Schneider},
  A.~{Shelden}, E.~S. {Sheldon}, A.~{Simmons}, S.~{Snedden}, J.~L. {Tinker},
  M.~{Viel}, B.~A. {Weaver}, D.~H. {Weinberg}, M.~{White}, C.~{Y{\`e}che}, and
  D.~G. {York}, {\it {Baryon acoustic oscillations in the Ly{$\alpha$} forest
  of BOSS quasars}},  {\em \aap} {\bf 552} (Apr., 2013) A96,
  [\href{http://arxiv.org/abs/1211.2616}{{\tt arXiv:1211.2616}}].

\bibitem{slosar13}
A.~{Slosar}, V.~{Ir{\v s}i{\v c}}, D.~{Kirkby}, S.~{Bailey}, N.~G. {Busca},
  T.~{Delubac}, J.~{Rich}, {\'E}.~{Aubourg}, J.~E. {Bautista}, V.~{Bhardwaj},
  M.~{Blomqvist}, A.~S. {Bolton}, J.~{Bovy}, J.~{Brownstein}, B.~{Carithers},
  R.~A.~C. {Croft}, K.~S. {Dawson}, A.~{Font-Ribera}, J.-M. {Le Goff}, S.~{Ho},
  K.~{Honscheid}, K.-G. {Lee}, D.~{Margala}, P.~{McDonald}, B.~{Medolin},
  J.~{Miralda-Escud{\'e}}, A.~D. {Myers}, R.~C. {Nichol}, P.~{Noterdaeme},
  N.~{Palanque-Delabrouille}, I.~{P{\^a}ris}, P.~{Petitjean}, M.~M. {Pieri},
  Y.~{Pi{\v s}kur}, N.~A. {Roe}, N.~P. {Ross}, G.~{Rossi}, D.~J. {Schlegel},
  D.~P. {Schneider}, N.~{Suzuki}, E.~S. {Sheldon}, U.~{Seljak}, M.~{Viel},
  D.~H. {Weinberg}, and C.~{Y{\`e}che}, {\it {Measurement of baryon acoustic
  oscillations in the Lyman-{$\alpha$} forest fluctuations in BOSS data release
  9}},  {\em \jcap} {\bf 4} (Apr., 2013) 26,
  [\href{http://arxiv.org/abs/1301.3459}{{\tt arXiv:1301.3459}}].

\bibitem{palanque13}
N.~{Palanque-Delabrouille}, C.~{Y{\`e}che}, A.~{Borde}, J.-M. {Le Goff},
  G.~{Rossi}, M.~{Viel}, {\'E}.~{Aubourg}, S.~{Bailey}, J.~{Bautista},
  M.~{Blomqvist}, A.~{Bolton}, J.~S. {Bolton}, N.~G. {Busca}, B.~{Carithers},
  R.~A.~C. {Croft}, K.~S. {Dawson}, T.~{Delubac}, A.~{Font-Ribera}, S.~{Ho},
  D.~{Kirkby}, K.-G. {Lee}, D.~{Margala}, J.~{Miralda-Escud{\'e}}, D.~{Muna},
  A.~D. {Myers}, P.~{Noterdaeme}, I.~{P{\^a}ris}, P.~{Petitjean}, M.~M.
  {Pieri}, J.~{Rich}, E.~{Rollinde}, N.~P. {Ross}, D.~J. {Schlegel}, D.~P.
  {Schneider}, A.~{Slosar}, and D.~H. {Weinberg}, {\it {The one-dimensional
  Ly{$\alpha$} forest power spectrum from BOSS}},  {\em \aap} {\bf 559} (Nov.,
  2013) A85, [\href{http://arxiv.org/abs/1306.5896}{{\tt arXiv:1306.5896}}].

\bibitem{palanque15}
N.~{Palanque-Delabrouille}, C.~{Y{\`e}che}, J.~{Baur}, C.~{Magneville},
  G.~{Rossi}, J.~{Lesgourgues}, A.~{Borde}, E.~{Burtin}, J.-M. {LeGoff},
  J.~{Rich}, M.~{Viel}, and D.~{Weinberg}, {\it {Neutrino masses and cosmology
  with Lyman-alpha forest power spectrum}},  {\em \jcap} {\bf 11} (Nov., 2015)
  011, [\href{http://arxiv.org/abs/1506.05976}{{\tt arXiv:1506.05976}}].

\bibitem{bautista15}
J.~E. {Bautista}, S.~{Bailey}, A.~{Font-Ribera}, M.~M. {Pieri}, N.~G. {Busca},
  J.~{Miralda-Escud{\'e}}, N.~{Palanque-Delabrouille}, J.~{Rich}, K.~{Dawson},
  Y.~{Feng}, J.~{Ge}, S.~G.~A. {Gontcho}, S.~{Ho}, J.~M. {Le Goff},
  P.~{Noterdaeme}, I.~{P{\^a}ris}, G.~{Rossi}, and D.~{Schlegel}, {\it {Mock
  Quasar-Lyman-{$\alpha$} forest data-sets for the SDSS-III Baryon Oscillation
  Spectroscopic Survey}},  {\em \jcap} {\bf 5} (May, 2015) 060,
  [\href{http://arxiv.org/abs/1412.0658}{{\tt arXiv:1412.0658}}].

\bibitem{baur17}
J.~{Baur}, N.~{Palanque-Delabrouille}, C.~{Y{\`e}che}, A.~{Boyarsky},
  O.~{Ruchayskiy}, {\'E}.~{Armengaud}, and J.~{Lesgourgues}, {\it {Constraints
  from Ly-$\alpha$ forests on non-thermal dark matter including
  resonantly-produced sterile neutrinos}},  {\em ArXiv e-prints} (June, 2017)
  [\href{http://arxiv.org/abs/1706.03118}{{\tt arXiv:1706.03118}}].

\bibitem{bautista17}
J.~E. {Bautista}, N.~G. {Busca}, J.~{Guy}, J.~{Rich}, M.~{Blomqvist}, H.~{du
  Mas des Bourboux}, M.~M. {Pieri}, A.~{Font-Ribera}, S.~{Bailey},
  T.~{Delubac}, D.~{Kirkby}, J.-M. {Le Goff}, D.~{Margala}, A.~{Slosar}, J.~A.
  {Vazquez}, J.~R. {Brownstein}, K.~S. {Dawson}, D.~J. {Eisenstein},
  J.~{Miralda-Escud{\'e}}, P.~{Noterdaeme}, N.~{Palanque-Delabrouille},
  I.~{P{\^a}ris}, P.~{Petitjean}, N.~P. {Ross}, D.~P. {Schneider}, D.~H.
  {Weinberg}, and C.~{Y{\`e}che}, {\it {Measurement of baryon acoustic
  oscillation correlations at z = 2.3 with SDSS DR12 Ly{$\alpha$}-Forests}},
  {\em \aap} {\bf 603} (June, 2017) A12,
  [\href{http://arxiv.org/abs/1702.00176}{{\tt arXiv:1702.00176}}].

\bibitem{bourboux17}
H.~{du Mas des Bourboux}, J.-M. {Le Goff}, M.~{Blomqvist}, N.~G. {Busca},
  J.~{Guy}, J.~{Rich}, C.~{Y{\`e}che}, J.~E. {Bautista}, {\'E}.~{Burtin}, K.~S.
  {Dawson}, D.~J. {Eisenstein}, A.~{Font-Ribera}, D.~{Kirkby},
  J.~{Miralda-Escud{\'e}}, P.~{Noterdaeme}, I.~{P{\^a}ris}, P.~{Petitjean},
  I.~{P{\'e}rez-R{\`a}fols}, M.~M. {Pieri}, N.~P. {Ross}, D.~J. {Schlegel},
  D.~P. {Schneider}, A.~{Slosar}, D.~H. {Weinberg}, and P.~{Zarrouk}, {\it
  {Baryon acoustic oscillations from the complete SDSS-III Ly$\alpha$-quasar
  cross-correlation function at $z=2.4$}},  {\em ArXiv e-prints} (Aug., 2017)
  [\href{http://arxiv.org/abs/1708.02225}{{\tt arXiv:1708.02225}}].

\bibitem{schaye00}
J.~{Schaye}, T.~{Theuns}, M.~{Rauch}, G.~{Efstathiou}, and W.~L.~W. {Sargent},
  {\it {The thermal history of the intergalactic medium$^{*}$}},  {\em \mnras}
  {\bf 318} (Nov., 2000) 817--826,
  [\href{http://arxiv.org/abs/astro-ph/9912432}{{\tt astro-ph/9912432}}].

\bibitem{ricotti00}
M.~{Ricotti}, N.~Y. {Gnedin}, and J.~M. {Shull}, {\it {The Evolution of the
  Effective Equation of State of the Intergalactic Medium}},  {\em \apj} {\bf
  534} (May, 2000) 41--56, [\href{http://arxiv.org/abs/astro-ph/9906413}{{\tt
  astro-ph/9906413}}].

\bibitem{theuns00}
T.~{Theuns} and S.~{Zaroubi}, {\it {A wavelet analysis of the spectra of
  quasi-stellar objects}},  {\em \mnras} {\bf 317} (Oct., 2000) 989--995,
  [\href{http://arxiv.org/abs/astro-ph/0002172}{{\tt astro-ph/0002172}}].

\bibitem{viel06}
M.~{Viel} and M.~G. {Haehnelt}, {\it {Cosmological and astrophysical parameters
  from the Sloan Digital Sky Survey flux power spectrum and hydrodynamical
  simulations of the Lyman {$\alpha$} forest}},  {\em \mnras} {\bf 365} (Jan.,
  2006) 231--244, [\href{http://arxiv.org/abs/astro-ph/0508177}{{\tt
  astro-ph/0508177}}].

\bibitem{theuns02}
T.~{Theuns}, S.~{Zaroubi}, T.-S. {Kim}, P.~{Tzanavaris}, and R.~F. {Carswell},
  {\it {Temperature fluctuations in the intergalactic medium}},  {\em \mnras}
  {\bf 332} (May, 2002) 367--382,
  [\href{http://arxiv.org/abs/astro-ph/0110600}{{\tt astro-ph/0110600}}].

\bibitem{bolton08}
J.~S. {Bolton}, M.~{Viel}, T.-S. {Kim}, M.~G. {Haehnelt}, and R.~F. {Carswell},
  {\it {Possible evidence for an inverted temperature-density relation in the
  intergalactic medium from the flux distribution of the Ly{$\alpha$} forest}},
   {\em \mnras} {\bf 386} (May, 2008) 1131--1144,
  [\href{http://arxiv.org/abs/0711.2064}{{\tt arXiv:0711.2064}}].

\bibitem{lidz10}
A.~{Lidz}, C.-A. {Faucher-Gigu{\`e}re}, A.~{Dall'Aglio}, M.~{McQuinn},
  C.~{Fechner}, M.~{Zaldarriaga}, L.~{Hernquist}, and S.~{Dutta}, {\it {A
  Measurement of Small-scale Structure in the 2.2-4.2 Ly{$\alpha$} Forest}},
  {\em \apj} {\bf 718} (July, 2010) 199--230,
  [\href{http://arxiv.org/abs/0909.5210}{{\tt arXiv:0909.5210}}].

\bibitem{bolton10}
J.~S. {Bolton}, G.~D. {Becker}, J.~S.~B. {Wyithe}, M.~G. {Haehnelt}, and
  W.~L.~W. {Sargent}, {\it {A first direct measurement of the intergalactic
  medium temperature around a quasar at z = 6}},  {\em \mnras} {\bf 406} (July,
  2010) 612--625, [\href{http://arxiv.org/abs/1001.3415}{{\tt
  arXiv:1001.3415}}].

\bibitem{becker11}
G.~D. {Becker}, J.~S. {Bolton}, M.~G. {Haehnelt}, and W.~L.~W. {Sargent}, {\it
  {Detection of extended He II reionization in the temperature evolution of the
  intergalactic medium}},  {\em \mnras} {\bf 410} (Jan., 2011) 1096--1112,
  [\href{http://arxiv.org/abs/1008.2622}{{\tt arXiv:1008.2622}}].

\bibitem{garzilli12}
A.~{Garzilli}, J.~S. {Bolton}, T.-S. {Kim}, S.~{Leach}, and M.~{Viel}, {\it
  {The intergalactic medium thermal history at redshift z = 1.7-3.2 from the
  Ly{$\alpha$} forest: a comparison of measurements using wavelets and the flux
  distribution}},  {\em \mnras} {\bf 424} (Aug., 2012) 1723--1736,
  [\href{http://arxiv.org/abs/1202.3577}{{\tt arXiv:1202.3577}}].

\bibitem{rudie12}
G.~C. {Rudie}, C.~C. {Steidel}, and M.~{Pettini}, {\it {The Temperature-Density
  Relation in the Intergalactic Medium at Redshift langzrang = 2.4}},  {\em
  \apjl} {\bf 757} (Oct., 2012) L30,
  [\href{http://arxiv.org/abs/1209.0005}{{\tt arXiv:1209.0005}}].

\bibitem{lee14}
K.-G. {Lee}, J.~P. {Hennawi}, D.~N. {Spergel}, D.~H. {Weinberg}, D.~W. {Hogg},
  M.~{Viel}, J.~S. {Bolton}, S.~{Bailey}, M.~M. {Pieri}, W.~{Carithers}, D.~J.
  {Schlegel}, B.~{Lundgren}, N.~{Palanque-Delabrouille}, N.~{Suzuki}, D.~P.
  {Schneider}, and C.~{Yeche}, {\it {IGM Constraints from the SDSS-III/BOSS DR9
  Ly-alpha Forest Flux Probability Distribution Function}},  {\em ArXiv
  e-prints} (May, 2014) [\href{http://arxiv.org/abs/1405.1072}{{\tt
  arXiv:1405.1072}}].

\bibitem{boera14}
E.~{Boera}, M.~T. {Murphy}, G.~D. {Becker}, and J.~S. {Bolton}, {\it {The
  thermal history of the intergalactic medium down to redshift z = 1.5: a new
  curvature measurement}},  {\em \mnras} {\bf 441} (July, 2014) 1916--1933,
  [\href{http://arxiv.org/abs/1404.1083}{{\tt arXiv:1404.1083}}].

\bibitem{bolton14}
J.~S. {Bolton}, G.~D. {Becker}, M.~G. {Haehnelt}, and M.~{Viel}, {\it {A
  consistent determination of the temperature of the intergalactic medium at
  redshift z = 2.4}},  {\em \mnras} {\bf 438} (Mar., 2014) 2499--2507,
  [\href{http://arxiv.org/abs/1308.4411}{{\tt arXiv:1308.4411}}].

\bibitem{rorai17}
A.~{Rorai}, J.~F. {Hennawi}, J.~{O{\~n}orbe}, M.~{White}, J.~X. {Prochaska},
  G.~{Kulkarni}, M.~{Walther}, Z.~{Luki{\'c}}, and K.-G. {Lee}, {\it
  {Measurement of the small-scale structure of the intergalactic medium using
  close quasar pairs}},  {\em Science} {\bf 356} (Apr., 2017) 418--422,
  [\href{http://arxiv.org/abs/1704.08366}{{\tt arXiv:1704.08366}}].

\bibitem{hiss17}
H.~{Hiss}, M.~{Walther}, J.~F. {Hennawi}, J.~{O{\~n}orbe}, J.~M. {O'Meara}, and
  A.~{Rorai}, {\it {A New Measurement of the Temperature Density Relation of
  the IGM From Voigt Profile Fitting}},  {\em ArXiv e-prints} (Oct., 2017)
  [\href{http://arxiv.org/abs/1710.00700}{{\tt arXiv:1710.00700}}].

\bibitem{narayanan00}
V.~K. {Narayanan}, D.~N. {Spergel}, R.~{Dav{\'e}}, and C.-P. {Ma}, {\it
  {Constraints on the Mass of Warm Dark Matter Particles and the Shape of the
  Linear Power Spectrum fro\ m the Ly{$\alpha$} Forest}},  {\em \apjl} {\bf
  543} (Nov., 2000) L103--L106,
  [\href{http://arxiv.org/abs/astro-ph/0005095}{{\tt astro-ph/0005095}}].

\bibitem{viel05}
M.~{Viel}, J.~{Lesgourgues}, M.~G. {Haehnelt}, S.~{Matarrese}, and A.~{Riotto},
  {\it {Constraining warm dark matter candidates including sterile neutrinos
  and light gravitinos with WMAP and the Lyman-{$\alpha$} forest}},  {\em \prd}
  {\bf 71} (Mar., 2005) 063534,
  [\href{http://arxiv.org/abs/astro-ph/0501562}{{\tt astro-ph/0501562}}].

\bibitem{uros06}
U.~{Seljak}, A.~{Makarov}, P.~{McDonald}, and H.~{Trac}, {\it {Can Sterile
  Neutrinos Be the Dark Matter?}},  {\em Physical Review Letters} {\bf 97}
  (Nov., 2006) 191303, [\href{http://arxiv.org/abs/astro-ph/0602430}{{\tt
  astro-ph/0602430}}].

\bibitem{viel08}
M.~{Viel}, G.~D. {Becker}, J.~S. {Bolton}, M.~G. {Haehnelt}, M.~{Rauch}, and
  W.~L.~W. {Sargent}, {\it {How Cold Is Cold Dark Matter? Small-Scales
  Constraints from the Flux Power Spectrum of the High-Redshift
  Lyman-{$\alpha$} Forest}},  {\em Physical Review Letters} {\bf 100} (Feb.,
  2008) 041304, [\href{http://arxiv.org/abs/0709.0131}{{\tt arXiv:0709.0131}}].

\bibitem{bird10}
S.~{Bird}, H.~V. {Peiris}, M.~{Viel}, and L.~{Verde}, {\it {Minimally
  parametric power spectrum reconstruction from the Lyman {$\alpha$} forest}},
  {\em \mnras} {\bf 413} (May, 2011) 1717--1728,
  [\href{http://arxiv.org/abs/1010.1519}{{\tt arXiv:1010.1519}}].

\bibitem{viel13WDM}
M.~{Viel}, G.~D. {Becker}, J.~S. {Bolton}, and M.~G. {Haehnelt}, {\it {Warm
  dark matter as a solution to the small scale crisis: New constraints from
  high redshift Lyman-{$\alpha$} forest data}},  {\em \prd} {\bf 88} (Aug.,
  2013) 043502, [\href{http://arxiv.org/abs/1306.2314}{{\tt arXiv:1306.2314}}].

\bibitem{baur15}
J.~{Baur}, N.~{Palanque-Delabrouille}, C.~{Y{\`e}che}, C.~{Magneville}, and
  M.~{Viel}, {\it {Lyman-alpha forests cool warm dark matter}},  {\em \jcap}
  {\bf 8} (Aug., 2016) 012, [\href{http://arxiv.org/abs/1512.01981}{{\tt
  arXiv:1512.01981}}].

\bibitem{yeche17}
C.~{Yeche}, N.~{Palanque-Delabrouille}, J.~{.~Baur}, and H.~{du Mas des
  BourBoux}, {\it {Constraints on neutrino masses from Lyman-alpha forest power
  spectrum with BOSS and XQ-100}},  {\em ArXiv e-prints} (Feb., 2017)
  [\href{http://arxiv.org/abs/1702.03314}{{\tt arXiv:1702.03314}}].

\bibitem{irsic17a}
V.~{Ir{\v s}i{\v c}}, M.~{Viel}, M.~G. {Haehnelt}, J.~S. {Bolton}, and G.~D.
  {Becker}, {\it {First Constraints on Fuzzy Dark Matter from Lyman-{$\alpha$}
  Forest Data and Hydrodynamical Simulations}},  {\em Physical Review Letters}
  {\bf 119} (July, 2017) 031302, [\href{http://arxiv.org/abs/1703.04683}{{\tt
  arXiv:1703.04683}}].

\bibitem{irsic17b}
V.~{Ir{\v s}i{\v c}}, M.~{Viel}, M.~G. {Haehnelt}, J.~S. {Bolton},
  S.~{Cristiani}, G.~D. {Becker}, V.~{D'Odorico}, G.~{Cupani}, T.-S. {Kim},
  T.~A.~M. {Berg}, S.~{L{\'o}pez}, S.~{Ellison}, L.~{Christensen}, K.~D.
  {Denney}, and G.~{Worseck}, {\it {New constraints on the free-streaming of
  warm dark matter from intermediate and small scale Lyman-{$\alpha$} forest
  data}},  {\em \prd} {\bf 96} (July, 2017) 023522,
  [\href{http://arxiv.org/abs/1702.01764}{{\tt arXiv:1702.01764}}].

\bibitem{armengaud17}
E.~{Armengaud}, N.~{Palanque-Delabrouille}, C.~{Y{\`e}che}, D.~J.~E. {Marsh},
  and J.~{Baur}, {\it {Constraining the mass of light bosonic dark matter using
  SDSS Lyman-$\alpha$ forest}},  {\em ArXiv e-prints} (Mar., 2017)
  [\href{http://arxiv.org/abs/1703.09126}{{\tt arXiv:1703.09126}}].

\bibitem{bolton16}
J.~S. {Bolton}, E.~{Puchwein}, D.~{Sijacki}, M.~G. {Haehnelt}, T.-S. {Kim},
  A.~{Meiksin}, J.~A. {Regan}, and M.~{Viel}, {\it {The Sherwood simulation
  suite: overview and data comparisons with the Lyman-alpha forest at redshifts
  $2 < z < 5$}},  {\em ArXiv e-prints} (May, 2016)
  [\href{http://arxiv.org/abs/1605.03462}{{\tt arXiv:1605.03462}}].

\bibitem{ma00}
C.-P. {Ma} and J.~N. {Fry}, {\it {Deriving the Nonlinear Cosmological Power
  Spectrum and Bispectrum from Analytic Dark Matter Halo Profiles and Mass
  Functions}},  {\em \apj} {\bf 543} (Nov., 2000) 503--513,
  [\href{http://arxiv.org/abs/astro-ph/0003343}{{\tt astro-ph/0003343}}].

\bibitem{seljak00}
U.~{Seljak}, {\it {Analytic model for galaxy and dark matter clustering}},
  {\em \mnras} {\bf 318} (Oct., 2000) 203--213,
  [\href{http://arxiv.org/abs/astro-ph/0001493}{{\tt astro-ph/0001493}}].

\bibitem{cooray02}
A.~{Cooray} and R.~{Sheth}, {\it {Halo models of large scale structure}},  {\em
  \physrep} {\bf 372} (Dec., 2002) 1--129,
  [\href{http://arxiv.org/abs/astro-ph/0206508}{{\tt astro-ph/0206508}}].

\bibitem{seljak09}
U.~{Seljak}, N.~{Hamaus}, and V.~{Desjacques}, {\it {How to Suppress the Shot
  Noise in Galaxy Surveys}},  {\em Physical Review Letters} {\bf 103} (Aug.,
  2009) 091303, [\href{http://arxiv.org/abs/0904.2963}{{\tt arXiv:0904.2963}}].

\bibitem{mead15}
A.~J. {Mead}, J.~A. {Peacock}, C.~{Heymans}, S.~{Joudaki}, and A.~F. {Heavens},
  {\it {An accurate halo model for fitting non-linear cosmological power
  spectra and baryonic feedback models}},  {\em \mnras} {\bf 454} (Dec., 2015)
  1958--1975, [\href{http://arxiv.org/abs/1505.07833}{{\tt arXiv:1505.07833}}].

\bibitem{baldauf13}
T.~{Baldauf}, U.~{Seljak}, R.~E. {Smith}, N.~{Hamaus}, and V.~{Desjacques},
  {\it {Halo stochasticity from exclusion and nonlinear clustering}},  {\em
  \prd} {\bf 88} (Oct., 2013) 083507,
  [\href{http://arxiv.org/abs/1305.2917}{{\tt arXiv:1305.2917}}].

\bibitem{ginzburg17}
D.~{Ginzburg}, V.~{Desjacques}, and K.~C. {Chan}, {\it {Shot noise and biased
  tracers: A new look at the halo model}},  {\em \prd} {\bf 96} (Oct., 2017)
  083528, [\href{http://arxiv.org/abs/1706.08738}{{\tt arXiv:1706.08738}}].

\bibitem{meiksin09}
A.~A. {Meiksin}, {\it {The physics of the intergalactic medium}},  {\em Reviews
  of Modern Physics} {\bf 81} (Oct., 2009) 1405--1469,
  [\href{http://arxiv.org/abs/0711.3358}{{\tt arXiv:0711.3358}}].

\bibitem{mcquinn15}
M.~{McQuinn}, {\it {The Evolution of the Intergalactic Medium}},  {\em ArXiv
  e-prints} (Nov., 2015) [\href{http://arxiv.org/abs/1512.00086}{{\tt
  arXiv:1512.00086}}].

\bibitem{miralda96}
J.~{Miralda-Escud{\'e}}, R.~{Cen}, J.~P. {Ostriker}, and M.~{Rauch}, {\it {The
  Ly alpha Forest from Gravitational Collapse in the Cold Dark Matter + Lambda
  Model}},  {\em \apj} {\bf 471} (Nov., 1996) 582,
  [\href{http://arxiv.org/abs/astro-ph/9511013}{{\tt astro-ph/9511013}}].

\bibitem{hernquist96}
L.~{Hernquist}, N.~{Katz}, D.~H. {Weinberg}, and J.~{Miralda-Escud{\'e}}, {\it
  {The Lyman-Alpha Forest in the Cold Dark Matter Model}},  {\em \apjl} {\bf
  457} (Feb., 1996) L51, [\href{http://arxiv.org/abs/astro-ph/9509105}{{\tt
  astro-ph/9509105}}].

\bibitem{prochaska10}
J.~X. {Prochaska}, J.~M. {O'Meara}, and G.~{Worseck}, {\it {A Definitive Survey
  for Lyman Limit Systems at z \~{} 3.5 with the Sloan Digital Sky Survey}},
  {\em \apj} {\bf 718} (July, 2010) 392--416,
  [\href{http://arxiv.org/abs/0912.0292}{{\tt arXiv:0912.0292}}].

\bibitem{sdss12gas}
P.~{Noterdaeme}, P.~{Petitjean}, W.~C. {Carithers}, I.~{P{\^a}ris},
  A.~{Font-Ribera}, S.~{Bailey}, E.~{Aubourg}, D.~{Bizyaev}, G.~{Ebelke},
  H.~{Finley}, J.~{Ge}, E.~{Malanushenko}, V.~{Malanushenko},
  J.~{Miralda-Escud{\'e}}, A.~D. {Myers}, D.~{Oravetz}, K.~{Pan}, M.~M.
  {Pieri}, N.~P. {Ross}, D.~P. {Schneider}, A.~{Simmons}, and D.~G. {York},
  {\it {Column density distribution and cosmological mass density of neutral
  gas: Sloan Digital Sky Survey-III Data Release 9}},  {\em \aap} {\bf 547}
  (Nov., 2012) L1, [\href{http://arxiv.org/abs/1210.1213}{{\tt
  arXiv:1210.1213}}].

\bibitem{schaye01}
J.~{Schaye}, {\it {Model-independent Insights into the Nature of the
  Ly{$\alpha$} Forest and the Distribution of Matter in the Universe}},  {\em
  \apj} {\bf 559} (Oct., 2001) 507--515,
  [\href{http://arxiv.org/abs/astro-ph/0104272}{{\tt astro-ph/0104272}}].

\bibitem{rees86}
M.~J. {Rees}, {\it {Lyman absorption lines in quasar spectra - Evidence for
  gravitationally-confined gas in dark minihaloes}},  {\em \mnras} {\bf 218}
  (Jan., 1986) 25P--30P.

\bibitem{bi97}
H.~{Bi} and A.~F. {Davidsen}, {\it {Evolution of Structure in the Intergalactic
  Medium and the Nature of the Ly{$\alpha$} Forest}},  {\em \apj} {\bf 479}
  (Apr., 1997) 523--542, [\href{http://arxiv.org/abs/astro-ph/9611062}{{\tt
  astro-ph/9611062}}].

\bibitem{font12}
A.~{Font-Ribera}, P.~{McDonald}, and J.~{Miralda-Escud{\'e}}, {\it {Generating
  mock data sets for large-scale Lyman-{$\alpha$} forest correlation
  measurements}},  {\em \jcap} {\bf 1} (Jan., 2012) 1,
  [\href{http://arxiv.org/abs/1108.5606}{{\tt arXiv:1108.5606}}].

\bibitem{press93}
W.~H. {Press}, G.~B. {Rybicki}, and D.~P. {Schneider}, {\it {Properties of
  high-redshift Lyman-alpha clouds. I - Statistical analysis of the
  Schneider-Schmidt-Gunn quasars}},  {\em \apj} {\bf 414} (Sept., 1993) 64--81,
  [\href{http://arxiv.org/abs/astro-ph/9303016}{{\tt astro-ph/9303016}}].

\bibitem{zuo94}
L.~{Zuo} and J.~R. {Bond}, {\it {The transmission correlation in the QSO
  Ly(alpha) forest produced by finite width lines}},  {\em \apj} {\bf 423}
  (Mar., 1994) 73--93.

\bibitem{liske98}
J.~{Liske}, J.~K. {Webb}, and R.~F. {Carswell}, {\it {Large-scale structure in
  the Lyman-alpha forest: a new technique}},  {\em \mnras} {\bf 301} (Dec.,
  1998) 787--796, [\href{http://arxiv.org/abs/astro-ph/9808082}{{\tt
  astro-ph/9808082}}].

\bibitem{1987MNRAS.227....1K}
N.~{Kaiser}, {\it {Clustering in real space and in redshift space}},  {\em
  \mnras} {\bf 227} (July, 1987) 1--21.

\bibitem{arinyo15}
A.~{Arinyo-i-Prats}, J.~{Miralda-Escud{\'e}}, M.~{Viel}, and R.~{Cen}, {\it
  {The non-linear power spectrum of the Lyman alpha forest}},  {\em \jcap} {\bf
  12} (Dec., 2015) 017, [\href{http://arxiv.org/abs/1506.04519}{{\tt
  arXiv:1506.04519}}].

\bibitem{seljak12}
U.~{Seljak}, {\it {Bias, redshift space distortions and primordial
  nongaussianity of nonlinear transformations: application to Ly-{$\alpha$}
  forest}},  {\em \jcap} {\bf 3} (Mar., 2012) 4,
  [\href{http://arxiv.org/abs/1201.0594}{{\tt arXiv:1201.0594}}].

\bibitem{cieplak15}
A.~M. {Cieplak} and A.~{Slosar}, {\it {Towards physics responsible for
  large-scale Lyman-{$\alpha$} forest bias parameters}},  {\em \jcap} {\bf 3}
  (Mar., 2016) 016, [\href{http://arxiv.org/abs/1509.07875}{{\tt
  arXiv:1509.07875}}].

\bibitem{kim04}
T.-S. {Kim}, M.~{Viel}, M.~G. {Haehnelt}, R.~F. {Carswell}, and S.~{Cristiani},
  {\it {The power spectrum of the flux distribution in the Lyman {$\alpha$}
  forest of a large sample of UVES QSO absorption spectra (LUQAS)}},  {\em
  \mnras} {\bf 347} (Jan., 2004) 355--366,
  [\href{http://arxiv.org/abs/astro-ph/0308103}{{\tt astro-ph/0308103}}].

\bibitem{vpfit}
B. Carswell et al.: http://www.ast.cam.ac.uk/$\sim$rfc/vpfit.html.

\bibitem{viel04DLA}
M.~{Viel}, M.~G. {Haehnelt}, R.~F. {Carswell}, and T.-S. {Kim}, {\it {The
  effect of (strong) discrete absorption systems on the Lyman {$\alpha$} forest
  flux power spectrum}},  {\em \mnras} {\bf 349} (Apr., 2004) L33--L37,
  [\href{http://arxiv.org/abs/astro-ph/0308078}{{\tt astro-ph/0308078}}].

\bibitem{font12b}
A.~{Font-Ribera} and J.~{Miralda-Escud{\'e}}, {\it {The effect of high column
  density systems on the measurement of the Lyman-{$\alpha$} forest correlation
  function}},  {\em \jcap} {\bf 7} (July, 2012) 28,
  [\href{http://arxiv.org/abs/1205.2018}{{\tt arXiv:1205.2018}}].

\bibitem{mcdonald05b}
P.~{McDonald}, U.~{Seljak}, R.~{Cen}, P.~{Bode}, and J.~P. {Ostriker}, {\it
  {Physical effects on the Ly{$\alpha$} forest flux power spectrum: damping
  wings, ionizing radiation fluctuations and galactic winds}},  {\em \mnras}
  {\bf 360} (July, 2005) 1471--1482,
  [\href{http://arxiv.org/abs/astro-ph/0407378}{{\tt astro-ph/0407378}}].

\bibitem{mcquinnwhite}
M.~{McQuinn} and M.~{White}, {\it {On estimating Ly{$\alpha$} forest
  correlations between multiple sightlines}},  {\em \mnras} {\bf 415} (Aug.,
  2011) 2257--2269, [\href{http://arxiv.org/abs/1102.1752}{{\tt
  arXiv:1102.1752}}].

\bibitem{2017arXiv170608532R}
K.~K. {Rogers}, S.~{Bird}, H.~V. {Peiris}, A.~{Pontzen}, A.~{Font-Ribera}, and
  B.~{Leistedt}, {\it {Simulating the effect of high column density absorbers
  on the one-dimensional Lyman-alpha forest flux power spectrum}},  {\em ArXiv
  e-prints} (June, 2017) [\href{http://arxiv.org/abs/1706.08532}{{\tt
  arXiv:1706.08532}}].

\bibitem{garzilli15b}
A.~{Garzilli}, T.~{Theuns}, and J.~{Schaye}, {\it {The broadening of
  Lyman-{$\alpha$} forest absorption lines}},  {\em \mnras} {\bf 450} (June,
  2015) 1465--1476, [\href{http://arxiv.org/abs/1502.05715}{{\tt
  arXiv:1502.05715}}].

\bibitem{mcquinn11}
M.~{McQuinn}, S.~P. {Oh}, and C.-A. {Faucher-Gigu{\`e}re}, {\it {On Lyman-limit
  Systems and the Evolution of the Intergalactic Ionizing Background}},  {\em
  \apj} {\bf 743} (Dec., 2011) 82, [\href{http://arxiv.org/abs/1101.1964}{{\tt
  arXiv:1101.1964}}].

\bibitem{altay11}
G.~{Altay}, T.~{Theuns}, J.~{Schaye}, N.~H.~M. {Crighton}, and C.~{Dalla
  Vecchia}, {\it {Through Thick and Thin -- H I Absorption in Cosmological
  Simulations}},  {\em \apjl} {\bf 737} (Aug., 2011) L37,
  [\href{http://arxiv.org/abs/1012.4014}{{\tt arXiv:1012.4014}}].

\bibitem{hui97}
L.~{Hui} and N.~Y. {Gnedin}, {\it {Equation of state of the photoionized
  intergalactic medium}},  {\em \mnras} {\bf 292} (Nov., 1997) 27,
  [\href{http://arxiv.org/abs/astro-ph/9612232}{{\tt astro-ph/9612232}}].

\bibitem{2016MNRAS.456...47M}
M.~{McQuinn} and P.~R. {Upton Sanderbeck}, {\it {On the intergalactic
  temperature-density relation}},  {\em \mnras} {\bf 456} (Feb., 2016) 47--54,
  [\href{http://arxiv.org/abs/1505.07875}{{\tt arXiv:1505.07875}}].

\bibitem{cen06}
R.~{Cen} and J.~P. {Ostriker}, {\it {Where Are the Baryons? II. Feedback
  Effects}},  {\em \apj} {\bf 650} (Oct., 2006) 560--572,
  [\href{http://arxiv.org/abs/astro-ph/0601008}{{\tt astro-ph/0601008}}].

\bibitem{gnedin98}
N.~Y. Gnedin and L.~Hui, {\it {Probing the universe with the Lyman alpha
  forest: 1. Hydrodynamics of the low density IGM}},  {\em Mon. Not. Roy.
  Astron. Soc.} {\bf 296} (1998) 44--55,
  [\href{http://arxiv.org/abs/astro-ph/9706219}{{\tt astro-ph/9706219}}].

\bibitem{kulkarni15}
G.~{Kulkarni}, J.~F. {Hennawi}, J.~{O{\~n}orbe}, A.~{Rorai}, and V.~{Springel},
  {\it {Characterizing the Pressure Smoothing Scale of the Intergalactic
  Medium}},  {\em \apj} {\bf 812} (Oct., 2015) 30,
  [\href{http://arxiv.org/abs/1504.00366}{{\tt arXiv:1504.00366}}].

\bibitem{2014MNRAS.444..503N}
Y.~{Noh} and M.~{McQuinn}, {\it {A physical understanding of how reionization
  suppresses accretion on to dwarf haloes}},  {\em \mnras} {\bf 444} (Oct.,
  2014) 503--514, [\href{http://arxiv.org/abs/1401.0737}{{\tt
  arXiv:1401.0737}}].

\bibitem{lai06}
K.~{Lai}, A.~{Lidz}, L.~{Hernquist}, and M.~{Zaldarriaga}, {\it {The Impact of
  Temperature Fluctuations on the Ly{$\alpha$} Forest Power Spectrum}},  {\em
  \apj} {\bf 644} (June, 2006) 61--70,
  [\href{http://arxiv.org/abs/astro-ph/0510841}{{\tt astro-ph/0510841}}].

\bibitem{1991ApJ...379..440B}
J.~R. {Bond}, S.~{Cole}, G.~{Efstathiou}, and N.~{Kaiser}, {\it {Excursion set
  mass functions for hierarchical Gaussian fluctuations}},  {\em \apj} {\bf
  379} (Oct., 1991) 440--460.

\bibitem{1996ApJ...462..563N}
J.~F. {Navarro}, C.~S. {Frenk}, and S.~D.~M. {White}, {\it {The Structure of
  Cold Dark Matter Halos}},  {\em \apj} {\bf 462} (May, 1996) 563,
  [\href{http://arxiv.org/abs/astro-ph/9508025}{{\tt astro-ph/9508025}}].

\bibitem{2017arXiv171106745D}
V.~{Desjacques}, D.~{Jeong}, and F.~{Schmidt}, {\it {Tidal shear and the
  consistency of microscopic Lagrangian halo approaches}},  {\em ArXiv
  e-prints} (Nov., 2017) [\href{http://arxiv.org/abs/1711.06745}{{\tt
  arXiv:1711.06745}}].

\bibitem{2003ApJ...596..768S}
J.~{Schaye}, A.~{Aguirre}, T.-S. {Kim}, T.~{Theuns}, M.~{Rauch}, and W.~L.~W.
  {Sargent}, {\it {Metallicity of the Intergalactic Medium Using Pixel
  Statistics. II. The Distribution of Metals as Traced by C IV}},  {\em \apj}
  {\bf 596} (Oct., 2003) 768--796,
  [\href{http://arxiv.org/abs/astro-ph/0306469}{{\tt astro-ph/0306469}}].

\bibitem{2004ApJ...602...38A}
A.~{Aguirre}, J.~{Schaye}, T.-S. {Kim}, T.~{Theuns}, M.~{Rauch}, and W.~L.~W.
  {Sargent}, {\it {Metallicity of the Intergalactic Medium Using Pixel
  Statistics. III. Silicon}},  {\em \apj} {\bf 602} (Feb., 2004) 38--50,
  [\href{http://arxiv.org/abs/astro-ph/0310664}{{\tt astro-ph/0310664}}].

\bibitem{2008ApJ...689..851A}
A.~{Aguirre}, C.~{Dow-Hygelund}, J.~{Schaye}, and T.~{Theuns}, {\it
  {Metallicity of the Intergalactic Medium Using Pixel Statistics. IV.
  Oxygen}},  {\em \apj} {\bf 689} (Dec., 2008) 851--864,
  [\href{http://arxiv.org/abs/0712.1239}{{\tt arXiv:0712.1239}}].

\bibitem{Garzilli15}
A.~{Garzilli}, A.~{Boyarsky}, and O.~{Ruchayskiy}, {\it {Cutoff in the Lyman
  $\{$$\backslash$alpha$\}$ forest power spectrum: warm IGM or warm dark
  matter?}},  {\em ArXiv e-prints} (Oct., 2015)
  [\href{http://arxiv.org/abs/1510.07006}{{\tt arXiv:1510.07006}}].

\bibitem{font12c}
A.~{Font-Ribera}, J.~{Miralda-Escud{\'e}}, E.~{Arnau}, B.~{Carithers}, K.-G.
  {Lee}, P.~{Noterdaeme}, I.~{P{\^a}ris}, P.~{Petitjean}, J.~{Rich},
  E.~{Rollinde}, N.~P. {Ross}, D.~P. {Schneider}, M.~{White}, and D.~G. {York},
  {\it {The large-scale cross-correlation of Damped Lyman alpha systems with
  the Lyman alpha forest: first measurements from BOSS}},  {\em \jcap} {\bf 11}
  (Nov., 2012) 059, [\href{http://arxiv.org/abs/1209.4596}{{\tt
  arXiv:1209.4596}}].

\bibitem{perez17}
I.~{P{\'e}rez-R{\`a}fols}, A.~{Font-Ribera}, J.~{Miralda-Escud{\'e}},
  M.~{Blomqvist}, S.~{Bird}, N.~{Busca}, H.~{du Mas des Bourboux},
  L.~{Mas-Ribas}, P.~{Noterdaeme}, P.~{Petitjean}, J.~{Rich}, and D.~P.
  {Schneider}, {\it {The SDSS-DR12 large-scale cross-correlation of damped
  Lyman alpha systems with the Lyman alpha forest}},  {\em \mnras} {\bf 473}
  (Jan., 2018) 3019--3038, [\href{http://arxiv.org/abs/1709.00889}{{\tt
  arXiv:1709.00889}}].

\end{thebibliography}\endgroup

\end{document}